\documentclass[a4paper]{ar2e}
\usepackage{amssymb}
\usepackage{amsmath}
\usepackage{natbib}
\usepackage{graphicx}
\def\aap{A{\&}A}
\def\pasp{Proc.Astr.Soc.Pacific}

\def\mnras{MNRAS}
\def\nat{Nature}
\def\apj{ApJ}

\def\apjl{ApJL}
\def\apjs{ApJS}

\def\comma{\;,}

\begin{document}


\input psfig.sty

\jname{..}
\jyear{2000}
\jvol{}
\ARinfo{1056-8700/97/0610-00}

\title{The inner regions of protoplanetary disks}

\markboth{The inner regions of protoplanetary disks}{The inner regions of protoplanetary disks}

\author{C. P. Dullemond
\affiliation{Max-Planck-Institute for Astronomy, 
K\"onigstuhl 17, D-69117 Heidelberg, Germany}
J. D. Monnier
\affiliation{University of Michigan, Astronomy Department, Ann Arbor, MI 48109 USA}
}

\begin{abstract}
  To understand how planetary systems form in the dusty disks around
  pre-main-sequence stars a detailed knowledge of the structure and
  evolution of these disks is required. While this is reasonably well
  understood for the regions of the disk beyond about 1 AU, the structure of
  these disks inward of 1 AU remains a puzzle. This is partly because it is
  very difficult to spatially resolve these regions with current telescopes.
  But it is also because the physics of this region, where the disk becomes
  so hot that the dust starts to evaporate, is poorly understood.  With
  infrared interferometry it has become possible in recent years to directly
  spatially resolve the inner AU of protoplanetary disks, albeit in a
  somewhat limited way. These observations have partly confirmed current
  models of these regions, but also posed new questions and
  puzzles. Moreover, it has turned out that the numerical modeling of these
  regions is extremely challenging. In this review we give a rough overview
  of the history and recent developments in this exciting field of 
  astrophysics.
\end{abstract}

\maketitle

\section{Introduction}
The dust- and gas-rich disks surrounding many pre-main sequence stars are of
great interest for gaining a better understanding of how planetary systems,
like our own, are formed. Since the first direct Hubble Space Telescope
images of such objects in silhouette against the background light in the
Orion Nebula \citep{McCaughrean:1996p271}, the observational and theoretical
study of these planetary birthplaces has experienced an enormous thrust,
leading to a much better understanding of what they are, what they look
like, how they evolve, how they are formed and how they are eventually
dissipated. While none of these aspects has yet been firmly understood in
detail, there is a consensus on the big picture. It is clear that
protoplanetary disks are the remnants of the star formation process. Excess
angular momentum of the original parent cloud has to be shed before most
matter can assemble into a ``tiny'' object such as a star. Protostellar
accretion disks \citep[see e.g.,][for a review]{hartmann:2009} are the most
natural medium by which this angular momentum can be extracted from the
infalling material. When the star is mostly ``finished'' and makes its way
toward the main sequence, the remainder of this protostellar accretion disk
is what constitutes a protoplanetary disk: the cradle of a future planetary
system.

\begin{figure}
\includegraphics[width=36em]{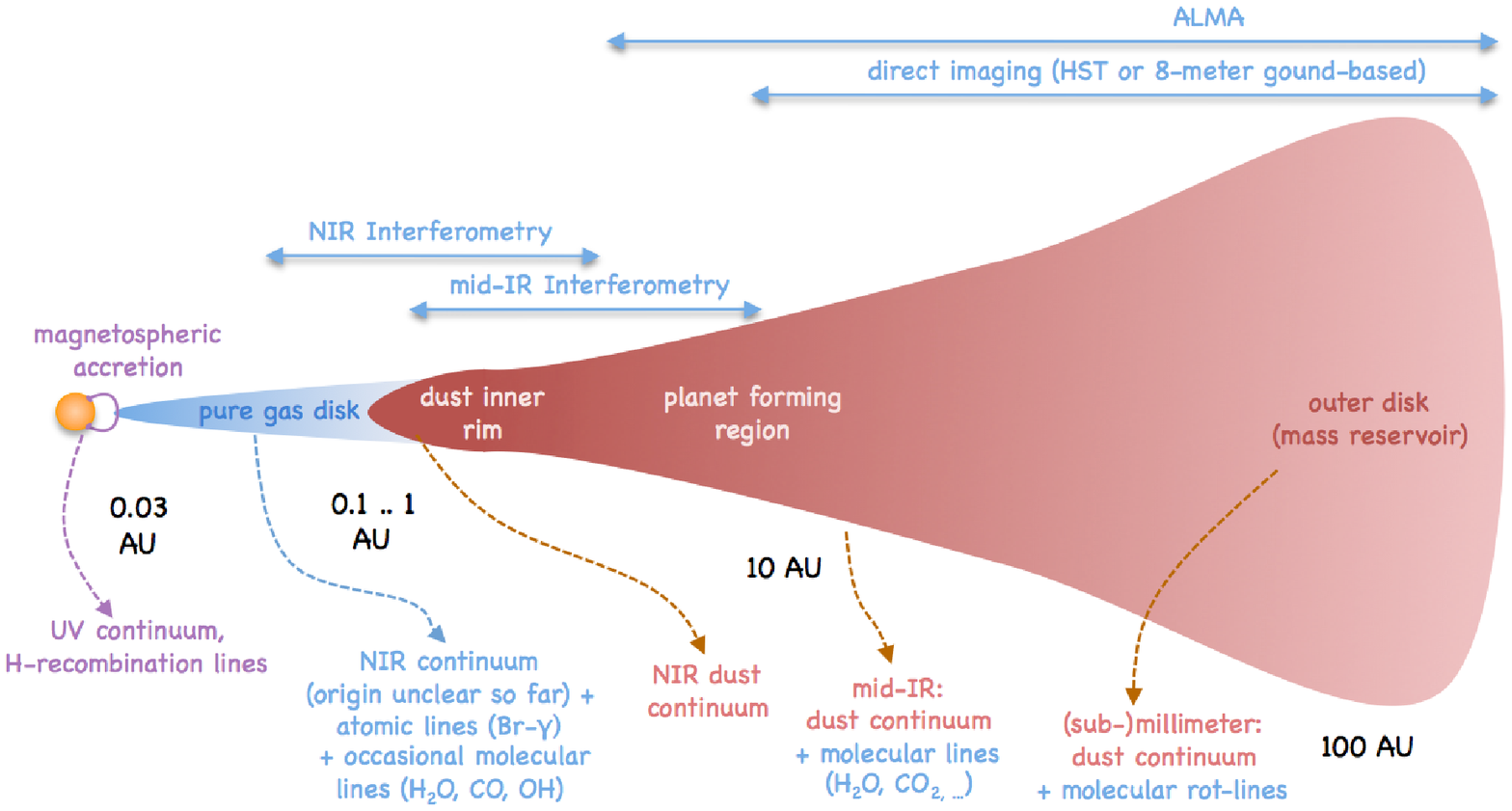}
\caption{\label{fig-picto-disk-scales}Pictogram of the structure and
spatial scales of a protoplanetary disk. Note that the radial scale on
the x-axis in not linear. Above the pictogram it is shown which techniques
can spatially resolve which scales. Below it is shown which kind of emission
arises from which parts of the disk.}
\end{figure}

Protoplanetary disks have a rich structure, with very different physics
playing a role in different regions of the disk. A pictographic
representation is shown in Fig.~\ref{fig-picto-disk-scales}. One can see the
strikingly large dynamic range that is involved: the outer radius of a
protoplanetary disk can be anywhere from a few tens of AU up to a 1000 AU or
more, while the inner disk radius is typically just a few stellar radii,
i.e.\ of the order of 0.02 AU. This spans a factor of $10^4\cdots 10^5$ in
spatial scale. For each orbit of the outer disk we have up to ten million
orbits of the inner edge of the disk. Equivalently, the dynamic time scale
on which various processes in the disk take place is also a million times
shorter (i.e.\ faster) in the very inner disk regions than in the outer disk
regions. And the temperatures differ also vastly: from $T\gg 10^3$ K in the
inner disk regions down to $T\sim 10\cdots 30$ K in the outer regions.

This large dynamic range in spatial scale, density and temperature means
that very different observational techniques have to be applied to probe the
various regions of these disks. Long wavelength telescopes (in the far
infrared and millimeter regime) predominantly probe the outer $\sim$ 10 -
few 100 AU of disks, while mid-infrared observations probe intermediate
radii $\sim$ few AU, near-infrared observations probe the inner AU and
finally optical and UV observations typically probe the regions very close
to the stellar surface ($\sim$ 0.01 - 0.1 AU). Of course, due to dust
scattering and sometimes due to quantum-heated grains, short wavelength
radiation can also be used to probe the outer disk regions, but for
thermally emitted radiation from the disk this rule of thumb applies very
well.

In this review we will limit our attention to the inner disk regions,
roughly inward of 1 AU. This is the region of the disk where temperatures
become high enough to modify, and even evaporate the dust of the disk. It is
also the region where a large amount of energy is set free in the form of
UV, optical and near infrared radiation, which plays a crucial role in the
overall energy balance of the disk.  Emission from this region is thus
easily observed. But until not long ago this region of the disk had been
inaccessible to spatially resolved observations because at typical distances
of 100 parsec they represent structures of 10 milli-arsecond scale, which is
clearly far smaller than what today's optical and infrared telescopes can
possibly resolve.  Spectroscopic observations have given many hints of
complex structure and interesting physics in these inner disk regions. With
spatially unresolved data alone there remains, however, a lot of ambiguity
in the physical interpretations of the observations.

With the advent of infrared interferometry this is now changing.  It has
become possible to directly spatially resolve these inner regions and verify
whether the conclusions drawn from spatially unresolved spectroscopic
observations in fact hold. The purpose of this review is to give an overview
of the current understanding of these inner disk regions, how this
understanding came about, and what are the many remaining open questions.
We will mainly focus on the transition region between the dusty outer disk
and the dust-free inner disk, the so called ``dust inner rim'' region,
because it is at these spatial scales that infrared interferometry has made
the largest impact. The topics of magnetospheric accretion
\citep[e.g.][]{Bouvier:2007p61976} and jet launching
\citep[e.g][]{Shang:2007p61866}, while extremely important in their own
right, will be considered as separate topics and will not be covered in this
review.

We will start our story with a rough historic review of how the conspicuous
near-infrared bump seen in the spectral energy distributions of many
protoplanetary disks led to the study of the dust-evaporation front in the
disk. Then we will discuss the advances in modeling this dust-gas transition
in the disk.  We will subsequently discuss the physics of the dust-free
inner disk, inward of the dust inner rim, and then discuss what is observed
from these very inner regions in terms of gas line emission. Finally we will
give a subjective outlook of remaining open issues and possible new areas of
investigation.

\section{Searching for the origin of the ``near-infrared bump''}
\subsection{The ``near-infrared bump'' and its first interpretations}
While the existence of circumstellar disks around high mass stars is still
very much debated as of this writing \citep{Cesaroni:2007p54847}, the
presence of such disks around young low mass pre-main sequence stars (T
Tauri stars and Brown Dwarfs) and their intermediate mass counterparts
(Herbig Ae/Be stars) has by now been firmly established
\citep{Watson:2007p321}. However, not all low/intermediate mass pre-main
sequence stars have disks. An often used indicator of the presence of a
circumstellar disk, or at least of circumstellar material, is infrared flux
in excess of what can possibly be explained by a stellar photosphere of a
reasonable size. By studying the fraction of stars with near infrared (NIR)
excess flux in young clusters of ages from 0.5 to 5 million years,
\citet{Haisch:2001p25755} could establish a clear trend: that the ``disk
fraction'' decreases with age, or in other words, that disks have a life
time of a few million years.

\begin{figure}
\includegraphics[width=36em]{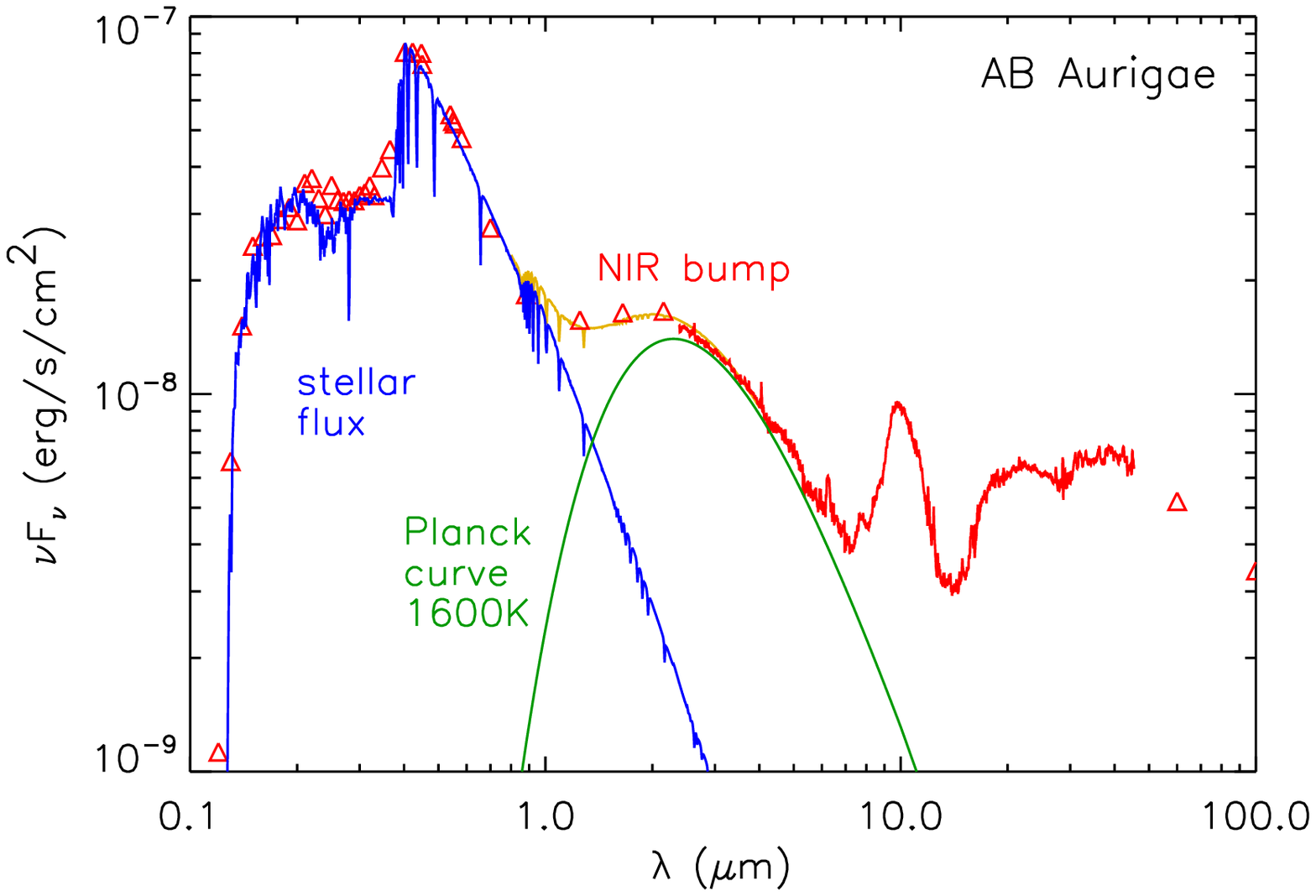}
\caption{\label{fig-ab-aur-sed}The spectral energy distribution of the
  Herbig Ae star AB Aurigae. Red is the measured emission. Blue is the
  steller spectrum predicted with a Kurucz stellar atmosphere model. The
  excess of flux above the atmosphere (the ``infrared excess'') is the
  thermal emission from the dust in the disk. The emission in the near
  infrared clearly has a bump-like structure, and is often called the ``near
  infrared bump''. In green a Planck curve at a temperature of 1600 K is
  overplotted. The golden curve is the sum of the Planck curve and the
  stellar atmosphere.}
\end{figure}

A question is, however, whether one can be sure that the NIR excess is
indeed from a disk and not from some circumstellar envelope or disk wind.
While we know from imaging that the cold outer circumstellar material is
indeed disk-like, little is known about the nature of the material inward of
what telescopes can spatially resolve.  The lack of correlation between
$A_V$ and NIR excess \citep{cohen:1979} is inconsistent with a spherical
dust geometry and the spectral shape of the infrared excess for T Tauri
stars and Brown Dwarfs can be explained fairly well with models of
irradiated dusty disks with a flat \citep{adams:1986} or flared shape
(Kenyon \& Hartmann \citeyear{Kenyon:1987p52281}; Calvet et
al.~\citeyear{calvet:1992}; Chiang \& Goldreich \citeyear{chiang:1997};
Menshchikov \& Henning \citeyear{Menshchikov:1997p61252}; D'Alessio et
al.~\citeyear{DAlessio:1998p71}). However, for Herbig Ae/Be stars this was
initially not so clear, and still remains under debate. It appears that the
JHKL photometric points nicely line up to form a ``bump'' very similar,
though not identical, to the peak of the Planck function at a temperature of
about $\sim$1500 K. This is perhaps most clearly seen in the spectrum of the
prototype Herbig Ae star AB Aurigae (Fig.~\ref{fig-ab-aur-sed}).
This NIR bump was not at all expected from any of the above mentioned models: They
tend to yield relatively smooth ``multi-color blackbody curves'' in which a
continuous series of Planck peaks at different temperatures add up to a
smooth curve. Now there appeared to be a single Planck peak in the spectrum,
albeit often with a bit of excess emission toward longer wavelengths. This ``NIR
bump'', as it is often called, is not just a small feature: it contains a
large amount of energy. The bump alone can contain up to half the infrared
flux from the entire system, and nearly all the emission originating from
the inner AU or so. It can therefore not be ignored: it must be understood
in terms of some physical model.

\citet{hillenbrand:1992} studied the NIR excesses of Herbig Ae/Be stars in a
systematic fashion. They interpreted this bump as originating from the hot
emission from an accretion disk. But to fit the NIR bump they had to make a
large inner hole in the disk, with a radius of about 0.1 AU (for AB~Aur,
$\sim$50$L_{\odot}$), inward of which there is no emission at all. And they
had to assume a relatively large accretion rate ($\dot M\sim
10^{-5}M_{\odot}/$yr) to match the flux levels of the NIR bump. Since in
standard accretion disk theory the hottest Planck component originates close
to the inner radius of the disk, this means that this inner radius was set
such that the temperature close to this inner radius was about 1500 K, the
temperature of the NIR bump. While this fits the NIR bump, a new question
emerges: If the inner edge of the accretion disk lies at $\sim$5-15 stellar
radii distance from the star, where does the accreting matter go after it
passes inward through this inner edge? If it would somehow continue inward,
then the inevitable release of accretional energy in these inner regions
would have to create hotter radiation in addition to the 1500 K emission,
for the reason of energy conservation. It would thus destroy the bump-like
shape of the NIR flux by filling in emission at shorter wavelength
\citep{hartmann:1993}.  Alternatively, if a magnetospheric accretion
scenario were responsible for this inner hole, the UV emission created by
the material flowing along these field lines and crashing onto the stellar
surface should be very strong, $L_{\mathrm{accr,magn}}\simeq 2GM_{*}\dot
M/R_{*}$. No such strong UV flux, consistent with accretion rates of the
order of $\dot M\sim 10^{-5}M_{\odot}/$yr, is observed. The only alternative
is that matter is simply ejected from this inner disk edge, but it is
unclear how this happens, and why this inner hole is at 1500
K in nearly all sources, not at different temperatures from source to
source. The Hillenbrand et al.\ study therefore rules out that the bump
originates from accretion alone.

\begin{figure}
\centerline{\includegraphics[width=30em]{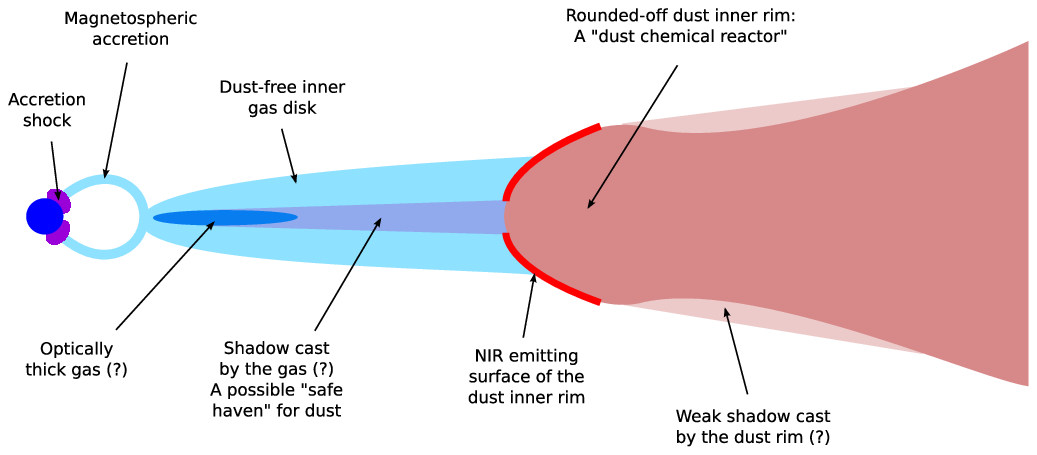}}
\caption{\label{fig-picto-inner-wall}Pictographic representation of the
  inner disk region out to a few AU.  One sees the magnetospheric accretion
  depicted near the star, the dust-free gas disk in the middle and the dust
  rim on the right.}
\end{figure}

So why would the NIR bump always be around 1500 K? To those who are familiar
with dust physics this number immediately calls to mind the process of dust
sublimation, often called ``dust evaporation'' in the astrophysical
community. This possible association was already pointed out in the
previously mentioned \citet{hillenbrand:1992} paper. Most species of
interstellar dust can survive as a solid only up to about that temperature,
with some rarer dust species surviving perhaps up to 1800 K. So it is a
reasonable idea to propose that the NIR emitting region around Herbig Ae/Be
stars consists of dust that is on the brink of evaporation. Since any model
of protoplanetary accretion disks predicts that the temperature of the disk
increases toward the star, it is quite reasonable to assume that inward of
some radius the disk is so hot that the dust simply evaporates. Since dust
is by several orders of magnitude the dominant carrier of continuum opacity
of any dust-gas mixture at ``low'' temperatures ($< 10^4$K), the
evaporation of dust in the very inner disk regions would render these disk
regions much less optically thick than the dusty parts of the disk, perhaps
even entirely optically thin. This means that the ``dust inner rim'', the
radial position separating the dusty outer regions from the dust-free inner
regions, should look like a kind of optically thick ``wall'' of dust at the
dust evaporation temperature. Inward of this wall the gas is either
optically thin, or at least much optically thinner than the dust, so that
this evaporation wall can be viewed by an observer right through the
transparent gas in front of it, as shown pictographically in
Fig.~\ref{fig-picto-inner-wall}.

This scenario was first proposed by \citet[][henceforth N01]{natta:2001}, in
their paper on the reinterpretation of Herbig Ae/Be spectral energy
distributions, and independently by \cite{tuthill:2001}, see Section
\ref{sec-nir-interfero-innerrim}. N01 clearly showed, using the disk model
of \citet[][henceforth C97]{chiang:1997}, that a disk model that does not
have such a dust evaporation wall can never fit the shape of the NIR bump,
but that such an inner dust wall naturally seems to explain it. It was
argued that this wall is ``puffed-up'' because it is much hotter than the
disk behind it, and thus is likely to have a much larger vertical scale
height. In a paper directly inspired by this work, \citet[][henceforth
DDN01]{dullemond:2001} subsequently integrated this puffed-up inner dust rim
model into the CG97 model and thus obtained a complete description of the
spectral energy distributions (SEDs) of Herbig Ae/Be stars in terms of a
simple irradiated disk model.

So if this simple model of the NIR bump is basically correct, then one may
wonder why mainly Herbig Ae/Be stars show such a huge bump while T Tauri
stars are not known for displaying such a conspicuous feature. DDN01 argue
that because the stellar luminosity is at much longer wavelengths for T
Tauri stars, a bump of this kind would be partly ``swamped'' by the flux
from the star, though a close look at the spectrum should still reveal such
a bump. On first sight T Tauri star SEDs do not show such a strong bump. But
through a careful subtraction of the stellar spectrum,
\citet{muzerolle:2003} show that also T Tauri stars consistently have such a
NIR bump, though perhaps weaker in relative sense than the Herbig stars. So
in that sense T Tauri stars are no different from Herbig stars.

\begin{figure}
\includegraphics[width=36em]{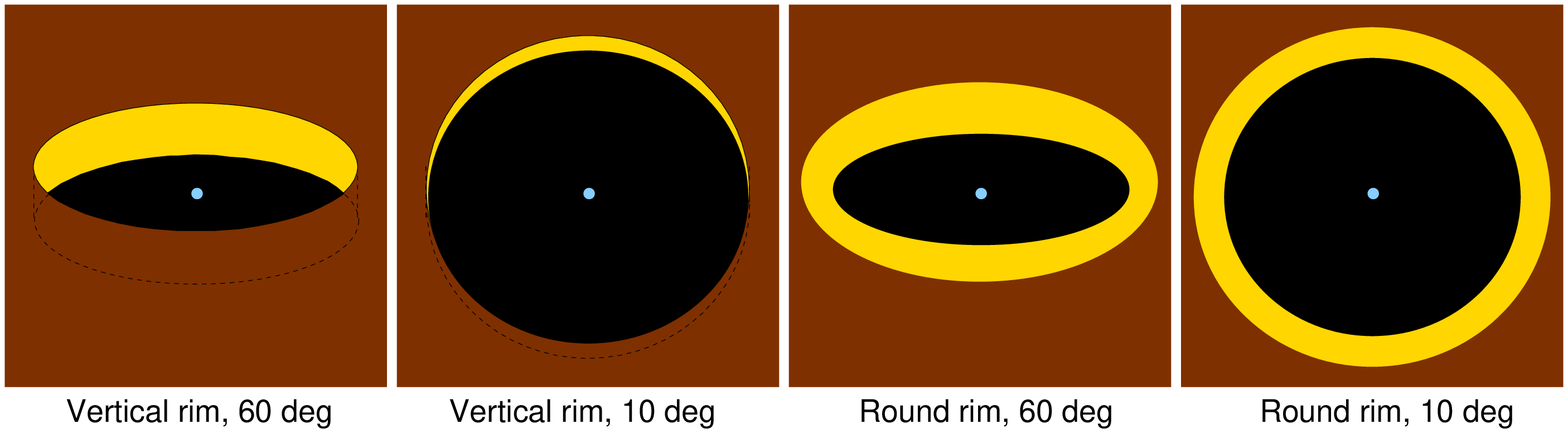}
\caption{\label{fig-viewing-the-rim}Pictographic representation of how an
  inner rim is viewed. The left two panels show a simple vertical rim model
  while the right two panels show a rounded-off rim. In both cases the left
  panel shows the rim at an inclination of about 60 degrees while the right
  panels shows it at near face-on inclination (10 degrees). The yellow color
  represents the emission from the hot dust that is in direct sight of the
  star and thus heated to high temperatures. Brown is the cooler dust behind
  the rim.}
\end{figure}

In spite of the early success of these models, there was no easy way of
telling with just NIR photometric data whether it was indeed the true nature
of these objects. Indeed, much simpler spherically symmetric envelope models
in which the dust was also removed inward of the dust evaporation radius,
could also fit the NIR bump and even in a number of cases the entire
spectral energy distribution \citep{pezzuto:1997, malfait:1998,
  miroshnichenko:1999, bouwman:2000, vinkovic:2006}. In fact, such models
appear to be more consistent with the lack of clear observed correlation
between the NIR flux and the disk inclination. For a simple perfectly
vertical wall model of the rim such a correlation is clearly expected, with
little NIR flux observed at near-face-on inclinations as illustrated in the
left two panels of Fig.~\ref{fig-viewing-the-rim}, and discussed in more
detail in Section \ref{sec-early-vertical-wall-models}.  Perhaps the most
clear counter-example is AB Aurigae, which has a huge NIR bump (see
Fig.~\ref{fig-ab-aur-sed}), but is known to be not very far from face-on
\citep[e.g.][]{Eisner:2003p50944,corder:2005}. Note, however, that AB
Aurigae is an object that is still surrounded by a substantial amount of
non-disk-related circumstellar material, which may contribute to the
NIR flux.

However, the key to distinguishing these models from each other is to
spatially resolve the NIR disk emission.  Since the spatial scale we are
talking about here is about 1 AU in diameter, which means 7 milliarcseconds
at typical distances of Herbig Ae stars, no NIR telescope is even remotely
able to make spatially resolved images of these structures to tell which
model is correct.

\subsection{Near-infrared interferometry: spatially resolving the inner disk regions}
\label{sec-nir-interfero-innerrim}
\begin{figure}
\includegraphics[width=4in]{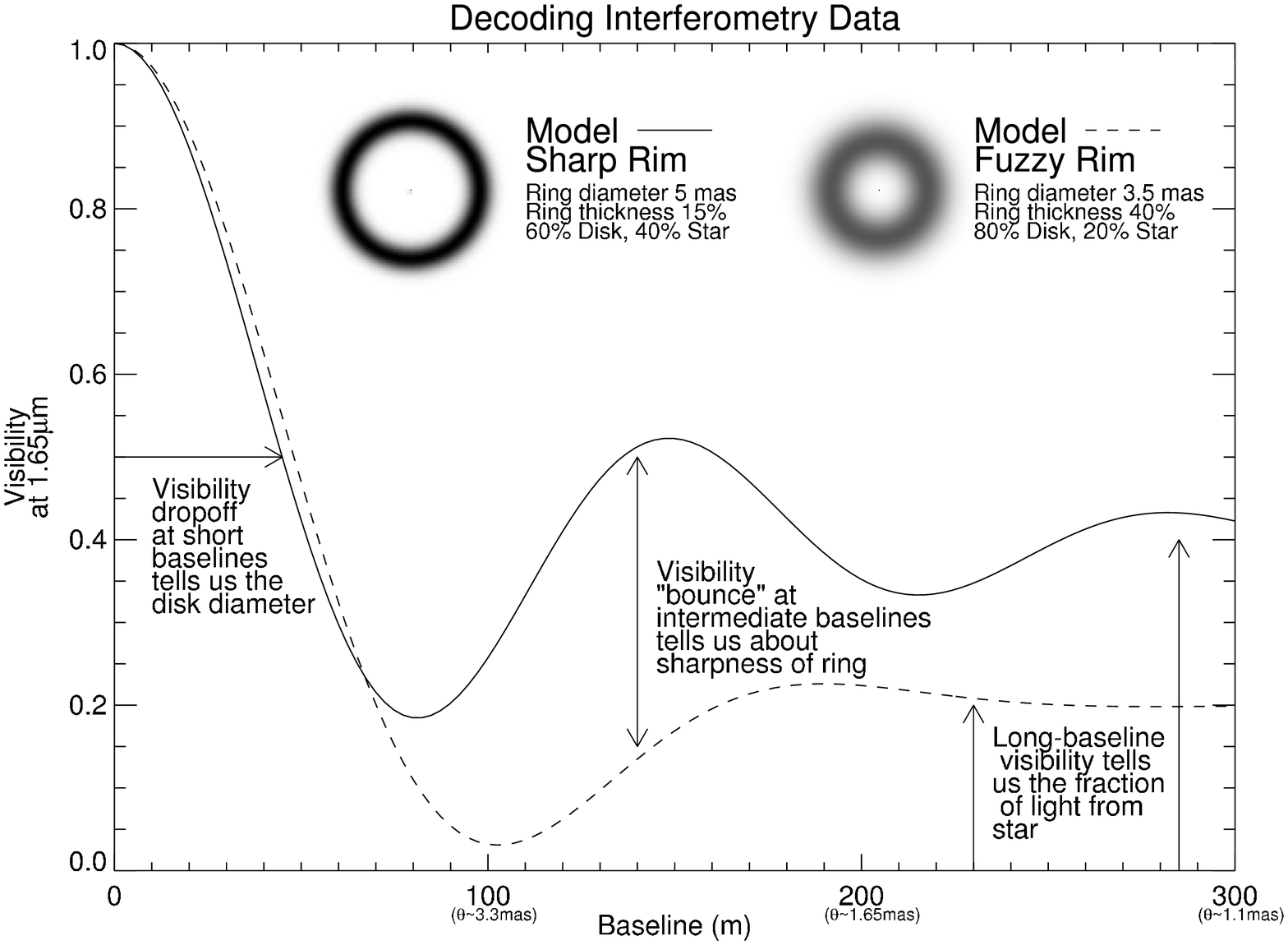}
\caption{
\label{fig-interfero}
{\em caption for box inset:}
}
\fbox{
\parbox{5in}{
\scriptsize
Single-baseline infrared interferometers have revolutionized our
understanding of the inner disk of young stellar objects by measuring
the size scale and morphology of the dust and gas at the inner edge
within an AU of the central star.
In this figure, we present a primer on how to decode interferometer
data as typically presented in observational papers.

The interferometer measures the
object's "Visibility" as a function of the telescopes separation ("Baseline"), with longer baselines probing finer angular scales.
Here we show two simple examples of inner disks,
a "Sharp Rim" model and a "Fuzzy Rim" Model
-- the intensity images are shown are included too.
Using arrows and labels, we show how measurements at different
baselines can be used to
directly constrain the (i) size of the disk, (ii) the sharpness of the
rim, and (iii) the fraction of light coming from the star and disk.
By combining multi-wavelength measurements from multiple
interferometers such as VLTI, Keck, and CHARA we can now span a wide
range of spatial scales necessary to unmask the true nature and
morphology of the inner regions of YSO disks.  New instruments are
being developed now to allow true imaging within the next few years.
}}
\end{figure}

Fortunately, at around the same time, a new observational technique started
to become mature: the technique of infrared interferometry.  This technique
allows to connect two or more infrared telescopes that are tens or hundreds
of meters apart, and thus achieve a spatial resolving power that far exceeds
that of a single telescope. The angular scale that can thus be probed equals
$\lambda/b$ radians, where $\lambda$ is the wavelength at which we are
observing and $b$ is the projected baseline, i.e.\ the distance between the
telescopes as projected on the sky toward the object we observe.  If we
observe at 2 $\mu$m with a projected baseline of 100 meter we arrive at
$2\times 10^{-8}$ radian, which is about 4 milliarcsecond. If our object is
at 140 parsec this leads to a spatial resolution of about 0.6 AU, which just
about matches what is needed to probe the dust evaporation front. So far,
however, the complexity of infrared interferometry has restricted the number
of telescopes that can participate in a single interferometric observation
to just two (a single baseline) or three (three baselines).  This means
that, in contrast to interferometry at radio wavelengths, infrared
interferometry does not yield actual aperture-synthesized images\footnote{Imaging of stellar surfaces has been successful \citep[e.g][]{monnier:2007a} with NIR interferometry just recently, but young star disks have been too faint for these techniques to date}. 
It is
limited to measuring ``sizes'', some information about radial brightness
profiles and ``asymmetries'' of emitting regions in the plane of the sky
(see Fig.~\ref{fig-interfero}). Yet, as we will see, this limited
information is enough to reveal a lot about the structure of the inner
regions of protoplanetary disks.

The star FU Orionis was the first young star to be targeted by infrared
interferometry \citep[using the Palomar Testbed Interferometer (PTI),][]{malbet:1998}. It is
the name-giving member of the class of ``FU Orionis stars'', a rare class of
young stellar objects with extremely high accretion rates \citep[$\dot M\sim
10^{-4} M_{\odot}$/yr;][]{hartmann:1985}, which are thus very bright in the
NIR. It was found that the measured ``size'' of the NIR emitting region of
this disk matched the expectations of classical accretion disk theory. 
This confirms the standard concept of FU Orionis
stars as being T Tauri stars with outbursting circumstellar disks.

Most T Tauri and Herbig Ae/Be stars, however, have much less active disks,
and it is with those stars that infrared interferometry allowed an
interesting discovery to be made. Using the Infrared Optical Telescope Array
(IOTA) interferometer, \citet{millan-gabet:1999a} found that the size of the
NIR emitting region for the ``normal'' Herbig Ae star AB~Aurigae is many
times larger than expected from the disk models current at that time. A
similar result was found for a T Tauri star by \citep{akeson:2000a}, and
further work confirmed these early conclusions \citep[especially the survey
of Herbig Ae disks by][]{millan-gabet:2001}.  At that time, before the
theoretical advances of the hot inner dust rim, the large measured sizes
were seen to support the spherical envelopes models
\citep[e.g.,][]{miroshnichenko:1999} which had the feature of a large,
optically-thin inner cavity.  Also, detailed measurements of AB~Aurigae
showed little size variation with position angle, consistent with a
spherical geometry.

\begin{figure}
\includegraphics[width=36em]{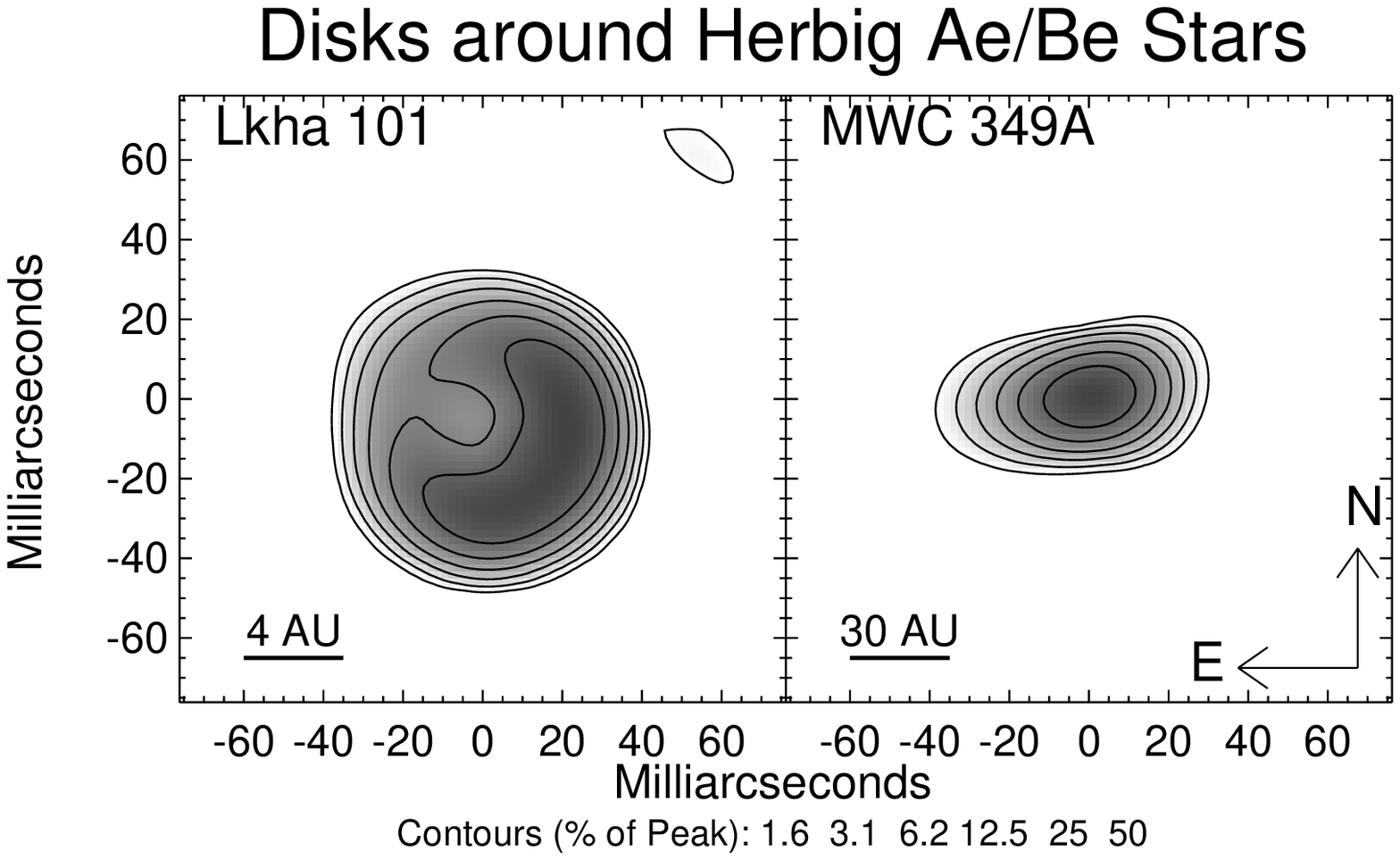}
\caption{\label{fig-lkha-101} These K-band images from Keck aperture masking
  showed the inner disks of the Herbig Be stars LkH$\alpha$~101
  \citep{tuthill:2001} and MWC~349 \citep{danchi:2001}. The dust-free inner
  cavity in LkH$\alpha$~101 was much larger than expected.}
\end{figure}

On the other hand, for two bright Herbig Be stars direct evidence for
disk-like (i.e.\ non-spherical) geometry was obtained using a variant of
infrared interferometry called ``aperture masking''. The aperture masking
technique uses a single telescope, in this case the Keck telescope
\citep{tuthill:2000}, and puts a mask in the optical beam to allow light
from only a partial set of sub-apertures to interfere at the camera focal
plane. This masking mimics a set of small telescopes observing
simultaneously as a miniature interferometer. Since the aperture mask can
contain relatively many sub-apertures, this technique allows for a true
image reconstruction, in contrast to ``long baseline'' infrared
interferometry.  With this technique \citet{tuthill:2001} presented a
spectacular image of the exceptionally bright Herbig Be star LkH$\alpha$
101, showing an asymmetric ring of dust emission around the central object
(Fig.~\ref{fig-lkha-101}).  This was interpreted by \citet{tuthill:2001} as
the inner dust rim of a flared circumstellar disk tilted slightly out of the
plane of the sky. The large size was consistent with a dust evaporation
radius only if the inner cavity was optically thin.  Subsequent Keck
aperture masking imaging of another Herbig Be star (MWC 349) also showed
clear disk-like geometry \citep{danchi:2001}. However, these two Herbig Be
sources remain the only two for which aperture masking interferometry could
be applied.

\subsection{Size-Luminosity Diagram}
\begin{figure}
\includegraphics[angle=90,width=36em]{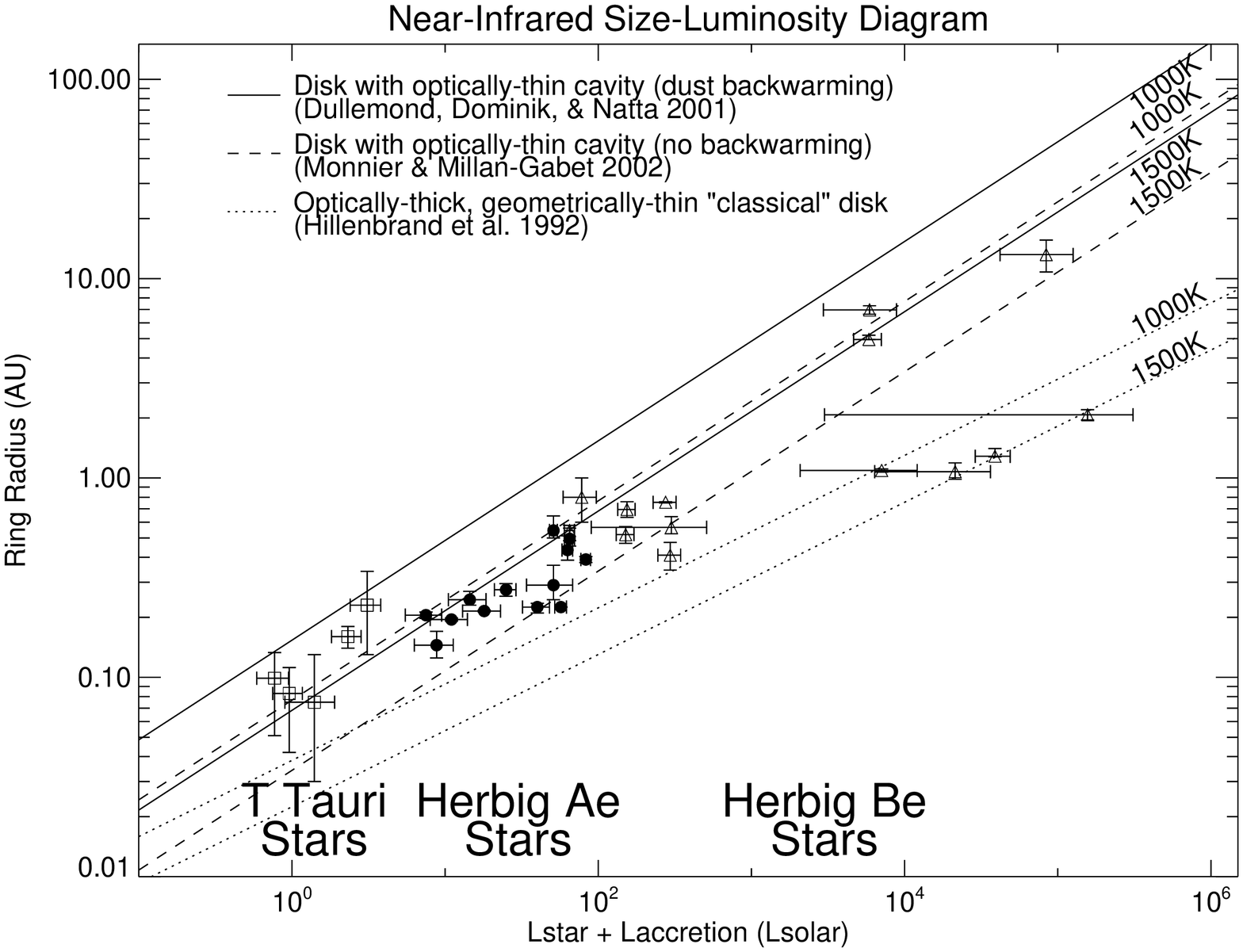}
\caption{\label{fig-scaling}The size-luminosity diagram obtained from NIR
  interferometric measurements of T Tauri and Herbig Ae stars \citep[adapted
  from figure presented in][]{millan-gabet:2007}.  The observed
  near-infrared sizes can be compared against different disk models,
  including disks with optically-thin cavities and those that are
  optically-thick but geometrically thin.  The most realistic disk models
  that include backwarming suggest dust evaporation temperatures are between
  1500--2000K.}
\end{figure}

Based on the disk geometry seen by aperture masking for Herbig Be stars, and
later confirmed for Herbig Ae disks using ``long baseline'' infrared
interferometry \citep{eisner:2003,eisner:2004}, we are led to consider disk
models to explain the interferometry size data.  Different classes of disk
models can be explored in a simple way using a ``Size-Luminosity diagram''.
To construct this diagram, the interferometer visibility data is fitted by
an emission ring model to represent the inner edge of the disk and this ring
radius is compared with the inferred central star luminosity from SED
fitting (Fig.~\ref{fig-scaling}); in this process, the fraction of light
coming from the star and disk must be estimated from an SED decomposition.
Monnier \& Millan-Gabet (\citeyear{Monnier:2002p52301}) showed that the
collective literature data of disk sizes could be best explained by an inner
rim of dust surrounding an optically-thin region, giving support to the
theoretical ideas of N01 and DDN01.  Covering four orders of magnitude in
luminosity, the measured sizes of the NIR emitting zone scale mostly as the
square root of the stellar luminosity, $R_{\mathrm{rim}}\propto
L_{*}^{1/2}$, as one would expect from the dust evaporation
radius. Moreover, these radii appear to be consistent with a dust
evaporation temperature between 1000-1500 K, the precise temperature values
depending on grain sizes and radiative transfer effects.  The initial
size-luminosity diagram, based on ``first-generation'' interferometry
measurements, also showed that some high luminosity sources (Herbig Be
stars) had smaller disk sizes that could be consistent with an
optically-thick inner disk, while (the very few measured) T Tauri disks
seemed broadly comparable to the Herbig Ae disks.

Following the first generation of measurements, larger samples of high
quality measurements were collected using the longest baseline
interferometers, especially the PTI and Keck Interferometers
\citep{eisner:2004,monnier:2005,akeson:2005a,eisner:2005}.  A summary of
these data are reproduced here in Figure~\ref{fig-scaling} \citep[originally
published in Protostars and Planets V;][]{millan-gabet:2007} and allow a detailed examination of
disk properties as a function of luminosity beyond the earlier work.  We see
here that with reasonable assumptions about disk backwarming (see Section
\ref{sec-dustrim}), the inferred dust evaporation temperatures are
typically between 1500-2000K even when assuming grey dust; as we shall
discuss in more detail later, these high temperatures are somewhat
problematic based on laboratory data of real grains.  Note that most of
these measurements were only done along one position angle of the disk; due
to projection effects, the true inner radius might be somewhat larger than
that measured.

Another key result from the size-luminosity diagram is a definite departure
from the $R_{\mathrm{rim}}\propto L_{*}^{1/2}$ scaling law for some (but not all!) of the
brightest sources in the sample, the Herbig B0-B3 stars: for these sources
the measured radii are smaller than the trend of the rest of the sample (see
Fig.~\ref{fig-scaling}).  In other words, they are undersized.  Apparently
the nature of the dust inner rim and/or the gas inward of the dust rim
changes for these very luminous sources.  \citet{monnier:2002} first
suggested that UV absorption by inner gas could explain undersized disks for
the higher luminosity targets.  The first serious study of this by
\citet{Eisner:2004p315} showed that the inner rim model appears to explain
the lower and intermediate luminosity sources, but that the high luminosity
sources appear to be better fit with a flat disk (i.e.\ an inner flat disk
of optically thick gas).

The early compendium of sizes showed some puzzling results for T Tauri
disks, too.  Some T Tauri disk sizes were quite large compared
to expectations based on the central star luminosities
\citep[e.g.,][]{colavita:2003,akeson:2005a, eisner:2005}.  In some cases,
the accretion luminosity of the infalling material is comparable to the
star's luminosity and must be included in the total central luminosity.
Given the larger errors bars, it seems the disk sizes seem to roughly obey
the same relations as Herbig Ae stars once the total central luminosity is
included \citep{muzerolle:2003,millan-gabet:2007}, but more work is needed
with higher angular resolution to improve the measurement precision.

\subsection{Beyond size-luminosity relations: Inner disk shapes}
\label{sec-inner-disk-shapes}
The size-luminosity diagram is a simplistic representation of an
increasingly large quantity of multi-wavelength, spatially-resolved
observations.  The diagram embeds a number of assumptions and can hide the
effect of important physical processes.  New data and modelling have
uncovered strong evidence for marked departures from the simple
``optically-thin'' inner cavity disk model with ``puffed-up'' inner wall.
The core set of observations are first described here and some of the
underlying physical mechanisms are explained and explored further in Section
\ref{sec-modeling}.

One of the predictions that the disk inner rim model makes is that the NIR
image on the sky should deviate from point-symmetry if the inner rim is seen
under an inclination angle that is sufficiently large. As can be clearly
seen in Fig.~\ref{fig-viewing-the-rim}, if the dust inner wall is perfectly
vertical then this deviation from point symmetry should be very strong: one
can only see the far side of the wall, not the near side. If the rim is very
much rounded off the situation softens a bit and point symmetry would only
be dramatically broken when looking at a relatively large inclination angle.
\citet{isella:2005} and \citet{tannirkulam:2007} both discovered very
natural reasons why the rim should be rounded off, as we shall discuss in
Section \ref{sec-modeling}.

The IOTA 3-telescope interferometer provided a chance to probe such asymmetries by
introducing the possibility of measuring a  quantity called the
``closure phase'' \citep{monnier:2007b}.  The closure phase is an observable
that is immune to atmospheric turbulence and gives precise information on the
disk geometry.
If an object is point-symmetric the measured closure phase should
be zero, while non-zero closure phase points to deviation from point
symmetry.  \citet{monnier:2006} reported that only few disks show a strong
non-zero closure phase, although half of their 16 sources did show small but
significant non-zero closure phase. An analysis showed that these
measurements rule out a vertical wall of dust, but are consistent with, and
provide support of, the idea of a rounded-off rim.

While models would remain simple if the near-IR emission only emerged from
the puffed-up rim, strong evidence is accumulating that there is an
additional even hotter component.  The first hints came from
\cite{eisner:2007a} who analyzed narrow-band infrared observations from PTI
interferometer and found that the disk sizes showed a small but systematic
wavelength dependence.  The disk emission region appears to be somewhat
smaller at shorter wavelengths than longer (2.0$\mu$m compared to
2.3$\mu$m), as would be encountered if there was a hot component
contributing to the emission inside the simplistic ``dust evaporation''
boundary.  Indeed, a strict prediction of a hot inner wall model is that the
disk emission sizes at different near-IR wavelengths should all be the same.
The idea of ``hot'' emission (maybe gas or refractory dust) inside the main
puffed-up inner wall is also suggested by multi-wavelength modeling of
interferometry data by \citet{isella:2008} and \citet{kraus:2008}.

The proposition of an extra emission component is now secure with recent
long-baseline ($\sim$300m) data from the CHARA interferometer: the inner
regions of the Herbig Ae prototypes MWC~275 and AB~Aur cannot be described
by just a simple puffed-up inner wall.  \citet{tannirkulam:2008b} reported two
major discoveries using the ``record-breaking'' 1.5~milliarcsecond resolution data from CHARA: the inner
edge of the disk is ``fuzzy'' and the amount of infrared excess is much
greater than expected from traditional SED decomposition into a star $+$ near-IR bump (see Figure~\ref{fig-ab-aur-sed}).

The ``fuzzy'' inner boundary is inferred from the lack of a visibility
``bounce'' from the long-baseline CHARA interferometer observations (see
Fig.~\ref{fig-interfero}) and proves that multiple components must exist in
the inner AU of these disks. The high NIR excess ($\sim$85\% at 2.2$\mu$m)
is probably impossible to generate from a passive disk that simply
re-radiates stellar luminosity (see Section \ref{sec-irexcess}), and points
importantly to a new energy source or to a radically different geometry.
These data also prove the scales on which the excess emission arises to be
well outside the star's surface and must be associated with the inner disk
(or possibly an inner disk wind?).  A more extensive dataset from VLTI on
MWC~275 (Benisty et al., in press) gives more details on the inner disk
morphology and multi-wavelength properties and also supports the idea that
the H and K band emission arises from a broad range of radii (``fuzzy'' rim)
instead of only a narrow range (``sharp'' rim).  Note that these data have
only been obtained for the brightest Herbig Ae stars and it is unclear
whether this ``fuzzy'' rim is a general feature of most YSO disks (including
T Tauri stars) or if this only applies to highly-accreting objects.

In addition to the rich spatial, multi-wavelength data, there are
other signs of dynamic and non-trivial physics operating in the inner
disk.  \citet{Wisniewski:2008p14444} and \citet{sitko:2008} report
significant changes in the environment of MWC~275 (see further
discusions in Section~\ref{sec-shadow-or-not}) that could be related
to the the production of the observed jet in this system
\citep{devine:2000}.  There have also been other reports of inner disk
asymmetries in AB~Aur \citep{millan-gabet:2006} and LkH$\alpha$~101
\citep{tuthill:2002}.

These recent findings seem to show that, while the inner dust rim picture
may be true in its essense, the simple assumption of a completely
transparent and non-emitting inner hole inside of a well-defined dust rim is
likely to be too naive.  The origin of the inner ``hot'' emission is still a
mystery and the theoretical considerations are outlined in Section
\ref{sec-inner-gas}.  It might come from accretional energy release in the
gas inward of the dust rim \citep{Akeson:2005p118}, or from refractory dust
that survives close to the star because it is being ``protected'' by the gas
\citep{Monnier:2005p20044}, or even from an inner disk wind/evelope. But so
far none of these scenarios are conclusive.

New spectroscopic and even spectro-interferometric observations hold immense
promise to reveal the origin of this mysterious ``hot'' inner component.
Before summarizing these recent results, we want to re-visit disk theory and
tie together all the physical processes we have just been discussing.
Following the development of the new ``standard disk model'' we will come
back and discuss the most recent results and future promising avenues of
study.

\section{Models of the inner dust rim}
\label{sec-modeling}
Motivated by the many exciting observations discussed in the previous
section, the presumed inner dust rims have been studied from a theoretical
modeling perspective by a number of authors. The main ingredient for this
modeling is radiative transfer: the calculation of how the radiation from
the star enters the disk, diffuses through it and thus determines the
temperature structure of the disk in the region of the dust rim.  This
problem has proven to be extremely complex, and even as of this writing not
nearly the last word has been said on this topic. Let us now ``get our hands
dirty'' by not only reviewing the theoretical models of the last decade, but
also by going through some of the math. We will start with the simplest
models, those of N01 and DDN01.

\subsection{Early ``vertical wall'' models of the dust inner rim}
\label{sec-early-vertical-wall-models}
The earliest models of the dust inner rim, those of N01 and DDN01, were
rather primitive. The aim was to get a first order understanding of
the origin of the NIR bump without going into detailed radiative transfer
modeling. From basic considerations, for instance by assuming a disk mass of
0.01 $M_\odot$, the optical depth of the dusty inner disk was expected to be
extremely high as long as the dust was present. So the dust inner rim was
treated as an optically thick vertical wall at a radius $R_{\mathrm{rim}}$
such that the dust in the wall has a temperature of 1500 K, the dust
evaporation temperature. Calculating the
temperature of a given wall is, strictly speaking, a challenging task.  But
for these simple initial models it was assumed that the wall radiates like a
blackbody, i.e.\ it emits a flux $F_{\mathrm{cool}}=\sigma
T_{\mathrm{rim}}^4$, where $\sigma$ is the Stefan-Boltzmann constant. This
should be compensated by the irradiation of the wall by the star. Here it
was assumed that the gas inward of the wall is transparent, so we get
$F_{\mathrm{heat}}=L_{*}/(4\pi R_{\mathrm{rim}}^2)$.  Equating the two
gives:
\begin{equation}\label{eq-radius-of-wall-bb}
R_{\mathrm{rim}} = \sqrt{\frac{L_{*}}{4\pi\sigma T_{\mathrm{rim}}^4}}
= R_{*}\left(\frac{T_{*}}{T_{\mathrm{rim}}}\right)^2
\end{equation}
where in the second identity we defined $T_{*}=(L_{*}/4\pi
R_{*}^2\sigma)^{1/4}$. For the parameters of AB Aurigae
($R_{*}=2.4\;R_\odot$, $T_{*}=10000\;$K) and for $T_{\mathrm{rim}}=1500\;$K we
obtain $R_{\mathrm{rim}}=0.5\;$AU. 

This tells only half of the story. The other half is to calculate the
surface height of the rim $H_{\mathrm{s,rim}}$, because together with the
radius it determines the so-called ``covering fraction'' of the rim, i.e.\
the fraction of the sky as seen by the star that is covered by the dust in
the inner dust rim.  This determines the fraction of the stellar light that
is captured by the rim and converted into NIR radiation, thus directly
determining the strength of the NIR flux. The concept of covering fraction
is extremely important for any circumstellar matter that reprocesses the
stellar radiation to other wavelengths. While a protoplanetary disk can in
principle also produce its own energy through the accretion process, for
most cases this is only a small amount compared to the luminosity of the
star. So we can safely assume that the disk, and its dust inner rim, is just
a ``passive'' object: it absorbs stellar radiation, and reemits this
radiation in the infrared. The total NIR emission from the inner rim is then
easily calculated:
\begin{equation}\label{eq-lum-covering-fraction}
L_{\mathrm{rim}} \simeq \omega L_{*}
\end{equation}
where $\omega$ is the covering fraction:
\begin{equation}\label{eq-def-covering-fraction}
\omega \simeq \frac{H_{\mathrm{s,rim}}}{R_{\mathrm{rim}}}
\end{equation}
where $H_{\mathrm{s,rim}}$ is the surface height of the rim, which we will
discuss in more detail below. In the above derivation we assume that all the
emitted infrared radiation from the rim escapes the system and none is
re-absorbed by the disk. This assumption is approximately valid as long as
$H_{\mathrm{s,rim}}\ll R_{\mathrm{rim}}$. For such a simple cylindrically
shaped rim the luminosity $L_{\mathrm{rim}}$ is very an-isotropic: The
observed flux for an observer at infinity at an inclination $i$ is roughly
\begin{equation}
F_{\nu,\mathrm{rim}} \simeq 4\sin i
\frac{R_{\mathrm{rim}}H_{\mathrm{s,rim}}}{d^2}B_\nu(T_{\mathrm{rim}})
\comma
\end{equation}
for inclinations $i$ small enough that the outer disk does not obscure the
view to the inner disk. Here $d$ is the distance to the observer and
$B_\nu(T)$ is the Planck function at frequency $\nu$ and temperature
$T$. Clearly $F_{\nu,\mathrm{rim}}$ from such a simple cylindrical rim model
is zero for $i=0$, i.e.\ for face-on view. As already stated above, this
simplification has been proven to be inconsistent with observations.
%
%
Also, the spectrum is assumed to be a perfect Planck curve, which is a major
simplification as well. 

But what is the value of $H_{\mathrm{s,rim}}$, i.e.\ How vertically extended
is the dust rim? The first thing to do is to make an estimate of the
vertical hydrostatic structure of the disk at $R=R_{\mathrm{rim}}$. Let us
assume that the temperature of the rim is independent of the vertical
coordinate $z$, measured upward from the midplane. Let us also assume that
the disk is geometrically thin, so that $z/R\simeq \cos\theta \simeq
\pi/2-\theta\ll 1$, where $\theta$ is the latitudinal coordinate of the
spherical coordinate system, being $\theta=0$ at the pole and $\theta=\pi/2$
at the equator. For a disk rotating with a Keplerian rotation speed
$\Omega(R)=\Omega_K(R)\equiv\sqrt{GM_{*}/R^3}$ (with $M_{*}$ the stellar
mass) each gram of disk material experiences a vertical force $f_z\simeq
-\Omega_K^2 z$. The equation of hydrostatic equilibrium $dP/dz=-\rho f_z$
then has the following solution:
\begin{equation}\label{eq-gaussian-density}
\rho_{\mathrm{gas}}(R,z) = \frac{\Sigma_{\mathrm{gas}}}{\sqrt{2\pi}H_{\mathrm{p}}}
\exp\left(-\frac{z^2}{2H_{\mathrm{p}}^2}\right)
\end{equation}
where $\Sigma_{\mathrm{gas}}$ is the surface density of the gas:
$\Sigma_{\mathrm{gas}}\equiv \int_{-\infty}^{+\infty}\rho(z)dz$ and the
pressure scale height $H_{\mathrm{p}}$ is given by
\begin{equation}\label{eq-hp-definition}
H_{\mathrm{p}} = \sqrt{\frac{k TR^3}{\mu_g GM_{*}}}
\end{equation}
where $\mu_g\simeq 2.3 m_p$ is the mean molecular weight, $m_p$ the proton
mass and $k$ the Boltzmann constant. Applying these formulae to
$R=R_{\mathrm{rim}}$ and $T=T_{\mathrm{evap}}$ yields $H_{\mathrm{p,rim}}$.
For AB Aurigae, our example star with a mass of $M_{*}\simeq 2.4\,
M_{\odot}$, and for $T_{\mathrm{evap}}=1500$ K we get $H_{\mathrm{p,rim}}
\simeq 0.036\;R_{\mathrm{rim}}$.

The pressure scale height of the disk at the dust rim, $H_{\mathrm{p,rim}}$,
is not necessarily the same as the surface height $H_{\mathrm{s,rim}}$. The
rim surface height is defined as the height above which stellar photons can
pass substantially beyond $R_{\mathrm{rim}}$ without being absorbed. To
compute this height we need to know the opacity of the dust in the rim as
seen by stellar photons. For small silicate dust grains a typical opacity in
the V band would be $\kappa_{\mathrm{sil}}(V)\simeq 10^4$
cm$^2$/gram-of-dust.  If we assume a gas-to-dust ratio of 100, then we can
estimate that stellar photons that move radially outward along the midplane
are absorbed in an extremely thin layer on the surface of the rim: a layer
defined by having an optical depth of about unity in the V band. The
geometric thickness of this layer is $\Delta
R=100/(\rho_{\mathrm{gas}}\kappa_{\mathrm{sil}})$ which for a gas surface
density of $\Sigma_{\mathrm{gas}}=100$ gram/cm$^2$ and a pressure scale height
of $H_{\mathrm{p,rim}}= 0.035 R_{\mathrm{rim}}$ amounts to $\Delta R\simeq
2.5\times 10^{-4}H_{\mathrm{p,rim}}\simeq 10^{-5}R_{\mathrm{rim}}$, indeed
an extremely thin layer.

Now, $\rho_{\mathrm{gas}}$ decreases with height
above the midplane. So a stellar photon moving radially outward at an angle
$\varphi$ with respect to the midplane will hit the inner dust wall at a
height $z\simeq \varphi R_{\mathrm{rim}}$, where according to
Eq.~(\ref{eq-gaussian-density}) the density is lower, and thus the geometric
thickness $\Delta R$ of the absorbing layer becomes larger.  At some height
$z$, $\Delta R$ becomes as large as a few times $H_{\mathrm{p,rim}}$ (in the
DDN01 paper it was taken to be 8 times): this is what we can define to be
the ``surface height'' of the rim, $H_{\mathrm{s,rim}}$. In practice
$H_{\mathrm{s,rim}}$ is of order 3 to 6 times $H_{\mathrm{p,rim}}$,
depending on the surface density $\Sigma_{\mathrm{gas}}$. 

It is important to note here that these are all rough estimates, and by no
means detailed calculations.  As we shall see shortly, the problem of
radiative transfer in the rim, i.e.\ the determination of the temperature
structure, and thus the vertical hydrostatic structure of the rim is a very
complex problem.

\subsection{Temperature of individual dust grains in the dust rim}
\label{sec-1d-trans-in-rim}
By assuming the rim to behave like a solid blackbody wall we have totally
ignored the individual particle character of the dust grains. Let us first
calculate the temperature of a single dust grain located at a distance $R$
from the star in a completely optically thin environment. If we assume that
the dust grain is a sphere of radius $a$ and that the optical absorption
cross section equals its geometric cross section $\pi a^2$ (amounting to a
``grey opacity'', i.e.\ an opacity independent of wavelength), then the
total amount of energy that the grain absorbs per second is $\pi a^2
L_{*}/4\pi R^2$.  If the grain has a temperature $T_{\mathrm{grey}}$, then
it emits blackbody radiation all over its surface, amounting to a cooling
rate of $4\pi a^2\sigma T^4_{\mathrm{grey}}$.  Equating the two yields:
\begin{equation}
T_{\mathrm{grey}} = \left(\frac{L_{*}}{16\pi \sigma R^2}\right)^{1/4}
= \left(\frac{R_{*}}{2R}\right)^{1/2}T_{*}
\end{equation}
If we put in the same numbers as we did before, for AB Aurigae, then we get
at $R=R_{\mathrm{rim}}\simeq 0.5\;$AU a grey dust temperature of
$T_{\mathrm{grey}}=1050\;$K, i.e.\ substantially smaller than the 1500 K we
calculated on the basis of the solid blackbody wall assumption.  So what has
gone wrong? One of the problems is that we did not include the
``backwarming'' of our dust grain by the thermal emission from the dust
further inside the wall. But a more precise analysis shows that we can only
do this properly if we solve the equations of radiative transfer in the disk
rim, as we shall do in Section \ref{sec-dustrim}. Moreover, our assumption
of a grey dust grain, i.e.\ a dust grain with a wavelength-independent
opacity, is likely to be wrong as well. 
\begin{figure}
\includegraphics[width=36em]{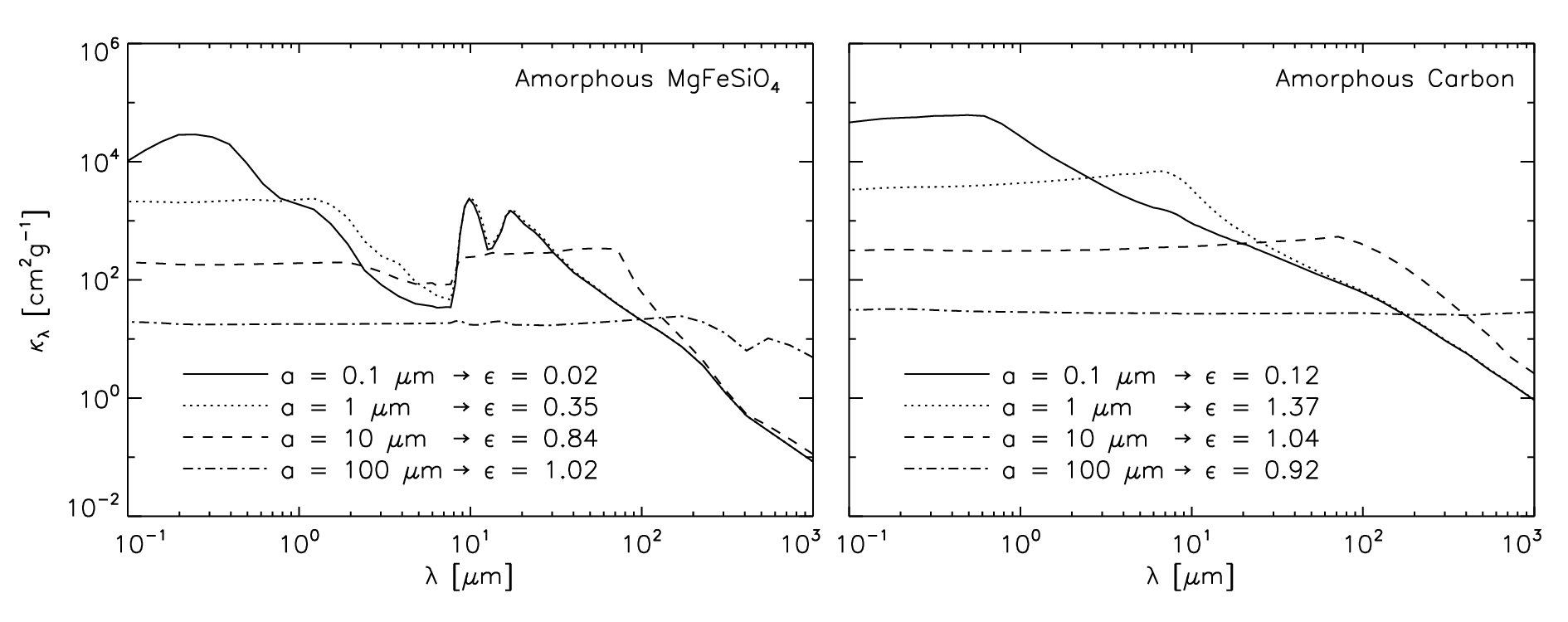}
\caption{\label{fig-dust-opacity}Absorption opacity per gram of dust for
  different dust grain sizes, assuming spherical compact grains. Left:
  amorphous olivine.  Right: amorphous carbon. The $\epsilon$ values
  according to Eq.~(\ref{eq-def-epsilon}) for a stellar blackbody
  temperature of $T_{*}=10000$K and a dust temperature of
  $T_{\mathrm{dust}}=1500$K are also given in the figure. Note that in
  reality in the inner rim the dust is presumably crystalline, but it is
  unclear what exact composition the dust will have then. For a review of
  astronomical dust and its opacities, see e.g.~\citet{Henning:2009p61359}
  and Henning {\bf (THIS ISSUE OF ARAA)}.}
\end{figure}
By accounting for the full opacity law $\kappa_\nu$ (see
Fig.~\ref{fig-dust-opacity}) we obtain the following equation for the dust
temperature of a single grain at distance $R$ from the star:
\begin{equation}\label{eq-dust-temp-full}
\int_0^{\infty} \kappa_\nu F_{*,\nu} d\nu = 4\pi \int_0^{\infty} \kappa_\nu B_{\nu}(T_{\mathrm{dust}}) d\nu 
\end{equation}
where $F_{*,\nu}=L_{*,\nu}/4\pi R^2=\pi (R_{*}/R)^2 B_\nu(T_{*})$ is the
flux from the star which we assume here, for convenience, to be a blackbody
emitter. Solving this equation for $T_{\mathrm{dust}}$ can be done
numerically. We can write Eq.~(\ref{eq-dust-temp-full}) in a way very
similar to the grey case, but with a correction factor:
\begin{equation}\label{eq-dust-temp-eps}
T_{\mathrm{dust}}=T_{*}\frac{1}{\epsilon^{1/4}}\sqrt{\frac{R_{*}}{2R}}
\end{equation}
where $\epsilon$ is defined as
\begin{equation}\label{eq-def-epsilon}
\epsilon \equiv \frac{\int_0^{\infty} \kappa_\nu B_{\nu}(T_{\mathrm{dust}}) d\nu/T_{\mathrm{dust}}^4}
{\int_0^{\infty} \kappa_\nu B_{\nu}(T_{*}) d\nu/T_{\mathrm{*}}^4}
\end{equation}
The symbol $\epsilon$ tells what is the ratio of effectiveness of emission
at wavelength at which the dust radiates away its heat and absorption at
stellar wavelengths. This ratio depends on the value of $T_{\mathrm{dust}}$
itself, so Eq.~(\ref{eq-dust-temp-eps}) is an implicit equation and requires
numerical iteration to solve (each time updating the $\epsilon$ value). But
it does tell us something about the radiation physics involved. For the case
of $\epsilon<1$ the stellar light (being at high temperatures and thus short
wavelengths) is absorbed at wavelengths where the opacity of the dust is
large, while the dust, being less than 1500 K in temperature, cools at
wavelengths where the opacity is smaller. This means that if $\epsilon<1$
one has $T_{\mathrm{dust}}>T_{\mathrm{grey}}$ while for $\epsilon>1$ one has
$T_{\mathrm{dust}}<T_{\mathrm{grey}}$. The ``efficiency factor'' $\epsilon$
therefore tells how efficiently the dust can cool. For $\kappa_\nu$
increasing with $\nu$, as is typical for small ($a\lesssim 1\mu$m) grains
(see Fig.~\ref{fig-dust-opacity}), we have $\epsilon<1$. Grains that are
substantially smaller than 1 micron in radius have a small efficiency
factor, $\epsilon\lesssim 0.5$, while grains larger than about ten microns
have an efficiency factor close to unity, $\epsilon\simeq 1$. This means
that if the dust rim consists of small grains only, we expect its inner
radius to lie further out than if the rim consists of large grains. This can
in fact be seen in the 2-D/3-D models described in Section
\ref{sec-2d-radtrans-in-rim}. In other words: large grains can survive
closer to the star than small grains.

Presumably crystalline grains can also survive closer to the star than
amorphous grains. In fact, dust at temperatures close to the sublimation
temperature is expected to be crystalline rather than amorphous as in
Fig.~\ref{fig-dust-opacity}. Pure crystals, even if they are very small,
will have a very low opacity in the V band and thus their ``efficiency
factor'' $\epsilon$ may be much larger than unity, allowing them to survive
close to the star. The question is, however, how pure such crystals are in
reality, and if they are in thermal contact with iron grains (which have a
very low $\epsilon$, hence high temperature).

\subsection{An approximate 1-D radiative transfer model of the dust rim}
\label{sec-dustrim}
Now let us come back to the issue of backwarming, i.e.\ the effect of mutual
heating of the grains by exchange of thermal radiation.  Solving the full
set of radiative transfer equations in a dust rim would require at least a
2-D treatment of the problem, as we shall discuss in Section
\ref{sec-2d-radtrans-in-rim}. This is a complex problem. But even a 1-D
horizontal approximation to the radiative transfer problem in the inner rim
is sufficiently complex that a full treatment requires a numerical
approach. Without going into any further detail, let us simply present a
formula for the dust temperature as a function of radial optical depth that
approximates the result of a full 1-D treatment reasonably well (Isella \&
Natta \citeyear{isella:2005}; D'Alessio et al.~\citeyear{DAlessio:2004p54841};
Muzerolle et al.~\citeyear{Muzerolle:2003p69};
Calvet et al.~\citeyear{Calvet:1991p8490}):
\begin{equation}\label{eq-milne-solution}
T_{\mathrm{dust}}(\tau_d) = T_{*} \sqrt{\frac{R_{*}}{2R}}
\left[\mu(2+3\mu\epsilon)+\left(\frac{1}{\epsilon}-3\epsilon\mu^2\right)
e^{-\tau_d/\mu\epsilon}\right]^{1/4}
\end{equation}
where $\tau_d$ is the optical depth as measured at wavelengths near the peak
of the Planck function for the dust temperature (for our inner rim analysis
this would be around 1500 K) and $\mu=\sin\phi$ with $\phi$ being the angle
under which the stellar radiation enters the dust wall. For stellar
radiation entering the wall near the equatorial plane, and thus
perpendicularly, one has $\phi=\pi/2$, i.e.\ $\mu=1$. On the other hand, if
one would choose $\mu=0$ and $\tau_d=0$, one recovers the equation for a
single dust grain without any nearby matter, Eq.~(\ref{eq-dust-temp-eps}).
The extra terms that appear for $\mu>0$ have to do with the backwarming
effect due to emission from other grains, and the $e^{-\tau_d/\mu\epsilon}$
factor has to do with the extinction of the direct stellar light as it
penetrates deeper into the rim. Note, by the way, that
Eq.~(\ref{eq-milne-solution}) assumes that $\epsilon=$constant, which is
of course just an approximation.

\begin{figure}
\includegraphics[width=35em]{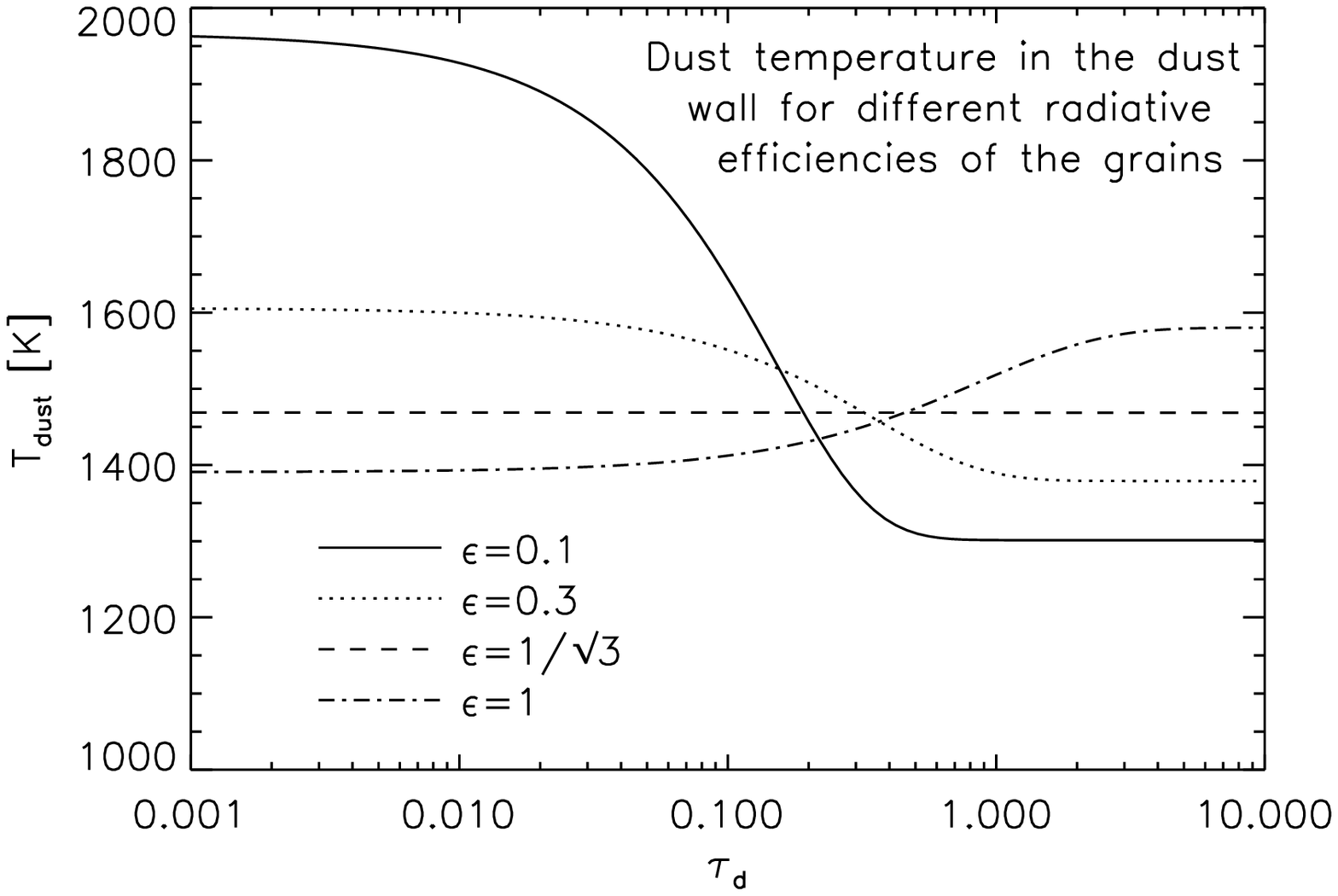}
\caption{\label{fig-1d-rt-inner-rim}The run of temperature with depth into
  the dust rim of our AB Aurigae example case, where $\tau_d$ is the optical
  depth in the NIR. The inner boundary of this model lies at 0.5 AU from the
  star. The stellar radiation enters the rim perpendicularly to the rim wall,
  i.e.\ $\mu=1$ in the equations in the text. The solution is shown for four
  different kinds of dust, i.e.\ four efficiency factors $\epsilon$. The
  case of $\epsilon=1$ corresponds typically to large dust grains ($a\gg
  3\mu$m) while smaller dust grains lead to smaller $\epsilon$. Note that
  the entire solution shown here corresponds only to a very thin layer at
  the inside of the dust wall.}
\end{figure}

Now let us plot the solutions for the parameters of AB Aurigae for different
values of $\epsilon$ (Fig.~\ref{fig-1d-rt-inner-rim}).  For the moment we
will ignore the issue of evaporation or condensation of dust.
The x-axis is plotted logarithmically, so the absolute inner edge of the
dust is at $\log(\tau_d)=-\infty$, i.e.\ infinitely much to the left of the
graph. For $\epsilon=1$ the dust near/at the inner edge (left part of the
graph) has a lower temperature than the dust deep inside the dust rim (right
part of the graph). The cause of this effect is a bit subtle, but it can be
traced back to the mutual radiative heating of the dust grains deep in the
rim: the infrared radiation that one grain emits can be absorbed by another
grain. This is the ``backwarming'' we mentioned earlier. A grain at
$\tau_d\gg 1$ is immersed in this radiation field, while for a grain at
$\tau_d\ll 1$ only half of the sky is delivering dust infrared emission. On
the other hand, that dust grain sees the stellar radiation field. For
$\epsilon=1$, however, the effect of mutual heating is stronger than the
direct stellar radiative heating, hence the larger temperature deeper in the
rim. A more detailed analysis would require some form of radiative transfer,
for instance the two-ray treatment explained in chapter 1 of the book by
Rybicki \& Lightman (\citeyear{Rybicki:1979p52302}).

The situation is reverse for the cases of $\epsilon=0.1$ and $\epsilon=0.3$
shown in the figure. There the dust at the very inner edge is quite hot
(1960 K if we ignore evaporation for the moment) while it is much cooler
deeper inside the rim (1300 K). This is because the dust that is in plain
sight of the star sees radiation that is very hot, and heats the grains at
wavelengths where $\kappa_\nu$ is much larger than at NIR wavelengths where
the grain cools. Heating is then more efficient than cooling, and thus the
dust becomes ``superheated''. Deep in the disk, however, there is no stellar
light available and this superheating effect is gone, hence the lower
temperature on the right side of the graph.

For a critical value of the efficiency parameter, $\epsilon=1/\sqrt{3}$, the
temperature is constant with $\tau_d$. Here all effects exactly cancel and
there is no temperature difference between the front and the back side of
the wall. This critical value of $\epsilon$ divides two very distinct kinds
of inner dust rim solutions, as was discovered by \citet{isella:2005}.
These solutions show that the dust rim is rarely a single-temperature
object, as was pointed out by \citet{Muzerolle:2003p69}.

\subsection{Is the rim a sharp, well-defined evaporation front?}
The reader may have already been bothered in the above radiative transfer
analysis by the fact that for all solutions except one there are regions
where the dust temperature exceeds the dust evaporation temperature of 1500
K. So what does this mean? For the low-$\epsilon$ solutions, with high
temperatures at the inner edge, this means that the star will simply
evaporate the dust and gradually ``eats its way'' outward into the disk,
moving the dust rim (and thus the above solution) outward until, through the
geometric $1/R^2$ dilution effect, the superheated dust temperature arrives
at the dust evaporation temperature. In other words: the inner rim of a disk
that consists of small $a\ll 3\mu$m grains is likely to lie at radii larger
than that predicted by Eq.~(\ref{eq-radius-of-wall-bb}). In fact, this
radius can be easily estimated, as long as $\epsilon\ll 1/\sqrt{3}$: under
those conditions the contribution of the backwarming by the grains deeper in
the rim becomes negligible and one can simply use
Eq.~(\ref{eq-dust-temp-eps}) and arrive at
\begin{equation}\label{eq-rwall-small-grains}
  R_{\mathrm{wall}} \simeq R_{*}\left(\frac{T_{*}}{T_{\mathrm{evap}}}\right)^2
  \frac{1}{2\sqrt{\epsilon}} = \frac{R_{\mathrm{wall,bb}}}{2\sqrt{\epsilon}}
\end{equation}
where $R_{\mathrm{wall,bb}}$ is the inner rim radius given by
Eq.~(\ref{eq-radius-of-wall-bb}). So the dust rim radius for small grains is
larger than that predicted by DDN01 if $\epsilon<0.25$. Right behind the new
location of the inner dust wall the temperature drops dramatically, and over
a very tiny spatial scale (remember the estimations of $\Delta R$ earlier).
The dust wall will thus have a flimsy layer of hot dust covering its inner
edge, but will be much cooler inside.

Now let us investigate what happens for $\epsilon>1/\sqrt{3}$. It appears
from Fig.~\ref{fig-1d-rt-inner-rim} that the grains on the inside of the
wall are below the evaporation temperature while those deep inside heat each
other up beyond the evaporation temperature. This gives rise to a very
confusing situation: all the dust deep inside the rim would evaporate while
the dust on the very inner edge would remain. This makes it impossible to
define a clear-cut location of the evaporation front. If we would include
2-D/3-D radiative transfer effects (see Section
\ref{sec-2d-radtrans-in-rim}) things would become a bit less extreme,
because radiation behind the dust rim can escape vertically and thus cool
down the regions some distance behind the rim. But we would still face the
confusing situation that we cannot make a clean definition of where
precisely the transition is between the hot inner dust-free disk and the
cooler outer dusty disk.

There are ideas floating in the literature on how Nature may solve this
apparent paradox. One is that instead of a sharply defined evaporation front
(a two-dimensional surface) we have what we could call a ``translucent
rim'': a three-dimensional volume of roughly optical depth unity, in which
the fraction of evaporated dust gradually increases from 0 to 1 as one moves
inward toward the star. This was predicted by \citet{isella:2005} and
\citet{Vinkovic:2006p55296} based on 1-D models, and something similar to
this was later indeed found in 2-D models by \citet{Kama:2009p62149}. For
the 2-D multi-grain size models of \citet{Tannirkulam:2007p46688}, however,
no such translucent rim was found. Progress in this area is, however, rather
slow because it appears evident that the solution of this problem lies in
the interplay between the complex physics of dust evaporation and
condensation with the numerically challenging problem of 2-D/3-D radiative
transfer (Section \ref{sec-2d-radtrans-in-rim}). It appears therefore that
the issue of how the evaporation front looks is still a wide open question
for future theoretical research.

\subsection{Better treatment of dust evaporation: a rounding-off of the rim}
\label{sec-isella-natta}
\begin{figure}
\includegraphics[width=35em]{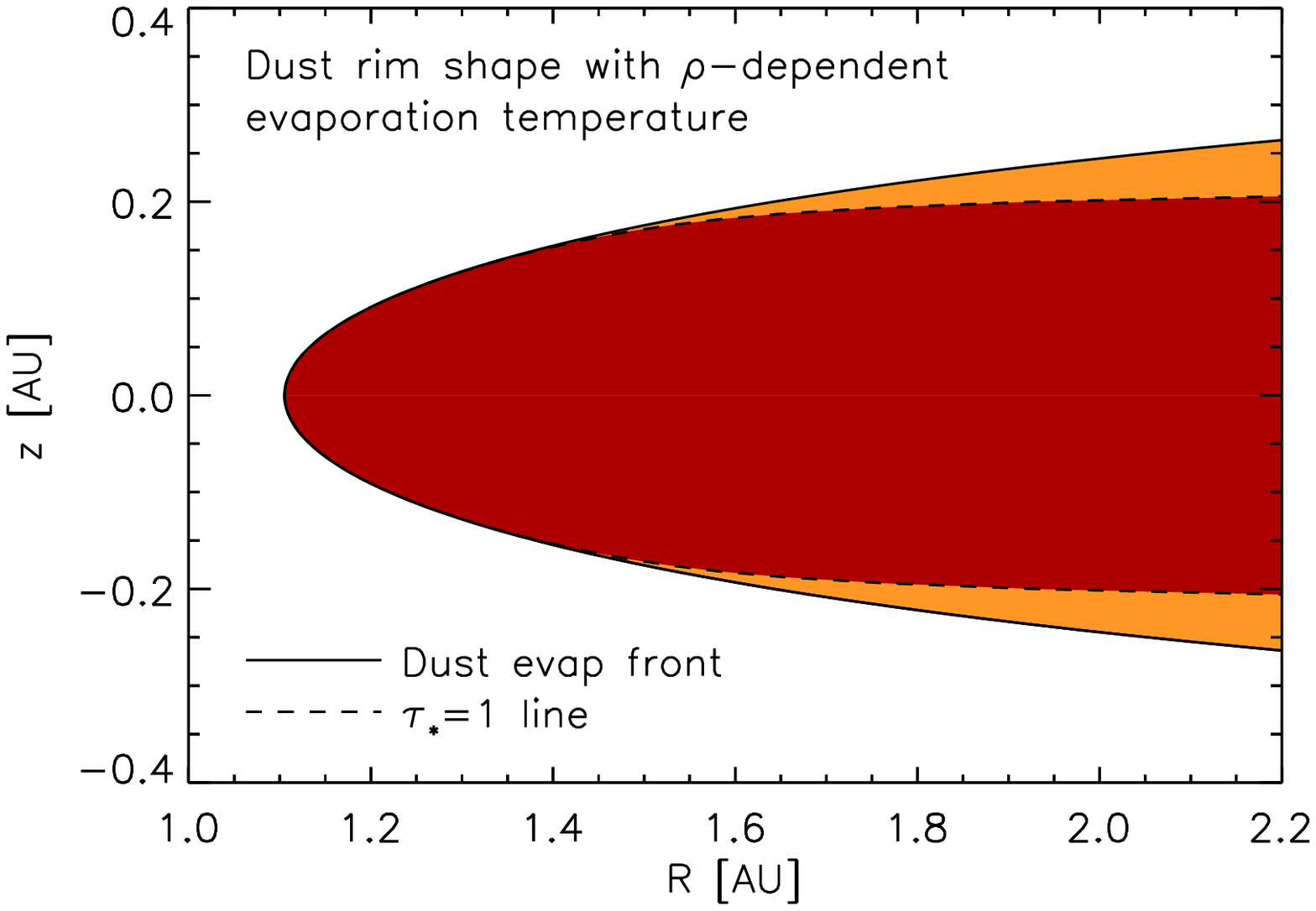}
\caption{\label{fig-rounded-rim-model}The rounded-off rim model of IN05
  computed using Eq.~(\ref{eq-rrim-in05-simple}), which is a simplified
  version of the IN05 model. Parameters are $R_{*}=2.4\;R_{\odot}$,
  $M_{*}=2.4\;M_\odot$, $T_{*}=10000$, $\epsilon=0.1$,
  $H_{\mathrm{p,rim}}=0.044\;$AU, $\Sigma_{\mathrm{gas}}=100\;$g/cm$^2$ and
  $\epsilon=0.1$. The brown region is the region that contains dust. The
  dark brown sub-region is the region that is more than $\tau_{*}=1$
  shielded from the stellar radiation field, where $\tau_{*}$ is defined as
  the radially outward optical depth at the average stellar wavelength. The
  midplane gas density at $\tau_{*}=1$ in this model is $\rho_g=6\times
  10^{-11}$g/cm$^3$.}
\end{figure}

So far, in our analysis of the dust rim, we always assumed a single
temperature at which the dust evaporates. This is, however, a simplification
with far reaching consequences, as \citet[henceforth IN05]{isella:2005}
found out. In reality, whether dust exists or not depends on a balance of
the processes of evaporation and condensation. While the evaporation rate
per unit area on grain surfaces depends on temperature only, the rate of
condensation depends also on the abundance of condensable atoms and
molecules in the gas phase. Typically, at a given temperature there is an
equilibrium partial pressure of vapor for which the evaporation and
condensation cancel each other out. If the partial pressure of the vapor is
below that value, any available solid state dust will continue to evaporate
until either the partial pressure of the vapor has reached the critical
vapor pressure or all the solids have been evaporated away. The process of
evaporation and condensation can be quite complex (Section
\ref{sec-behind-the-wall}), but if the abundances of condensable atoms and
molecules are known, it roughly amounts to having a temperature-dependent
critical total gas pressure $P_{\mathrm{gas,crit}}(T)$ above which dust can
exist in solid form and below which it cannot. In other words, for a given
gas density, there exists a critical temperature $T_{\mathrm{evap}}$ above
which the dust evaporates and below which it will condense. IN05 give a
fitting formula of the tabular results of \citet{Pollack:1994p209} which
reads: $T_{\mathrm{evap}} =G\rho_{\mathrm{gas}}^\gamma$ in which
$\rho_{\mathrm{gas}}$ is given in units of gram per cm$^3$ and the constants
are $G=2000$ and $\gamma=0.0195$. The dependence of $T_{\mathrm{evap}}$ on
density is weak, but since the density drops by a huge factor over a few
pressure scale heights (cf.~Eq.~\ref{eq-gaussian-density}) the evaporation
temperature drops considerably with height above the midplane. Adopting
small dust grains and thus using Eq.~(\ref{eq-rwall-small-grains}) as an
estimate of the evaporation radius as a function of $z$, and using
Eq.~(\ref{eq-gaussian-density}) for the vertical gas density distribution,
we arrive at the following shape of the disk rim:
\begin{equation}\label{eq-rrim-in05-simple}
R_{\mathrm{rim}}(z) 
\simeq R_{*}\frac{T_{*}^2}{2G^2\sqrt{\epsilon}} 
\left(\frac{\Sigma_{\mathrm{gas}}}{\sqrt{2\pi}H_{\mathrm{p,rim}}}\right)^{-2\gamma}
e^{\gamma z^2/H_{\mathrm{p,rim}}^2}
\end{equation}
We must make an estimation of $H_{\mathrm{p,rim}}$ and
$\Sigma_{\mathrm{gas}}$ even though the radius at which we should estimate
them would in principle depend on the outcome of the calculation itself.
For AB Aurigae, assuming $\Sigma_{\mathrm{gas}}=100$ gram/cm$^2$ and small
grains, say, grains with $\epsilon=0.1$, the midplane rim radius for this
model lies roughly around 1.0 AU. If we use the $\tau_d\rightarrow\infty$
temperature from Eq.~(\ref{eq-milne-solution}) for our chosen value of
$\epsilon$ (taking $\mu=1$), we get a temperature from which we can obtain,
through Eq.~(\ref{eq-hp-definition}), a reasonable estimate of the pressure
scale height $H_{\mathrm{p,rim}}$ behind the rim. In our example this
becomes $H_{\mathrm{p,rim}}=0.044\;$AU. With Eq.~(\ref{eq-rrim-in05-simple})
we can now calculate that the rim midplane radius is 1.1 AU, roughly confirming
our Ansatz of 1 AU. The resulting rim shape is plotted in
Fig.~\ref{fig-rounded-rim-model}.

The figure shows that the rounding-off of the rim is very strong. This is
not an effect that can be conveniently ignored. The figure also shows the
height above the midplane where the rim becomes optically thin (the dashed
line), showing that the rim can have $H_{\mathrm{s,rim}}$ as large as 0.2
AU, which in this case amounts to a covering fraction of about 20\%,
i.e.~producing an inclination-averaged NIR luminosity of
$L_{\mathrm{NIR}}\simeq 0.2\,L_{*}$. The dust evaporation temperature in the
rim ranges from 1270 K near the midplane down to 1050 K at the point where
the rim becomes optically thin. These dust evaporation temperatures are
smaller than the 1500 K assumed usually, and this may lead to the model bump
peaking at slightly longer wavelengths than observed (Isella et
al.~\citeyear{Isella:2006p48435}). But these results depend on the minerals
one assumes and on grain sizes. In fact, as we shall see in Section
\ref{sec-2d-radtrans-in-rim}, grain size distributions may also result in
the effect of rounding-off of the rim, in addition to the IN05 mechanism.

\subsection{Two-dimensional radiative transfer models}
\label{sec-2d-radtrans-in-rim}
To get deeper into the issue of the 2-D shape of the rim and the region just
behind it, it is clear that there is no path around utilizing 2-D continuum
radiative transfer codes. With ``2-D'' we actually mean 3-D, in the sense
the light is allowed to move in all 3 spatial directions. But the dust
temperatures and other quantities that result from these calculations have
axial symmetry: they only depend on the coordinates $R$ and $z$. In 
that sense these models are clearly 2-D. 

The problem of multi-dimensional radiative transfer can only be solved using
numerical methods. There are basically two main types of radiative transfer
algorithms in use today: those based on iterative integration of the formal
transfer equation along pre-defined photon directions (so called ``grid
based codes'' or better ``discrete ordinate codes'' because the angular
directions are modeled as an angular grid) and those based on a Monte Carlo
type of simulation of photon movement. In 1-D the discrete ordinate methods
have proven to be extremely efficient and usually superior over Monte Carlo
methods. But in 2-D and 3-D the discrete ordinate methods become cumbersome
and difficult to handle. Moreover, many of the advantages of these methods
are lost in 2-D and 3-D. Therefore, for problems of multi-dimensional
continuum radiative transfer in circumstellar disks and envelopes the Monte
Carlo method has become the most popular method. In particular the Monte
Carlo methods of Lucy (\citeyear{Lucy:1999p52313}) and of Bjorkman \& Wood
(\citeyear{Bjorkman:2001p469}) have proven to be ideally suited for this
problem, and by far most models today use these techniques or variants of
them. These methods are robust (they give reasonable answers even for
non-expert users) and easy to handle.  The main drawback is that in
very highly optically thick regions these codes can become slow
and/or give low photon statistics and hence high Monte Carlo
noise. Improvements in the efficiency of these methods, however, have made
this problem less serious in recent years (see e.g.\ Min et
al.~\citeyear{Min:2009p33666} and Pinte et al.~\citeyear{Pinte:2009p33678}).

The first 2-D frequency-dependent radiative transfer treatment of the inner
rim problem was presented, in the context of full disk models, by Dullemond
\& Dominik (\citeyear{Dullemond:2004p2858}, henceforth DD04a, though see
also Nomura \citeyear{Nomura:2002p49479}). These models gave new insight
into the actual appearance of the dust rim and the possible shadow it casts
on the disk behind it (see Sections \ref{sec-shadow-or-not} and
\ref{sec-irexcess}).

A main concern of the DD04a models is that they did not treat the
rounding-off of the rim. Unfortunately the 2-D radiative transfer treatment
of a rounded-off rim has proven to be a tremendous challenge, even
today. The reason is the extremely thin $\tau_{*}=1$ layer at the inner edge
of the rim (i.e.~$\Delta R\ll R_{\mathrm{rim}}$). For proper radiative
transfer one must spatially resolve the transition between optically thin
and optically thick.  This requires tremendously fine grid spacing at the
inner rim. For an inner rim that is aligned with the grid (a vertical rim
for cylindrical coordinates) this can be easily achieved by choosing an
$R$-grid that becomes finer and finer close to the rim, as is done in
DD04a. But a rounded-off rim is not aligned with the grid. This requires a
grid refinement that is not separable in $R$ and $z$, for instance by
allowing cells to be recursively split into 2$\times$2 (or in 3-D
2$\times$2$\times$2) subcells there where high spatial resolution is
required. Such a technique is often called ``adaptive mesh refinement''
(AMR). While some codes are technically able to do this, there are still
various numerical complications that arise with the dust evaporation on such
extremely refined grids.

\begin{figure}
\includegraphics[width=23em,angle=90]{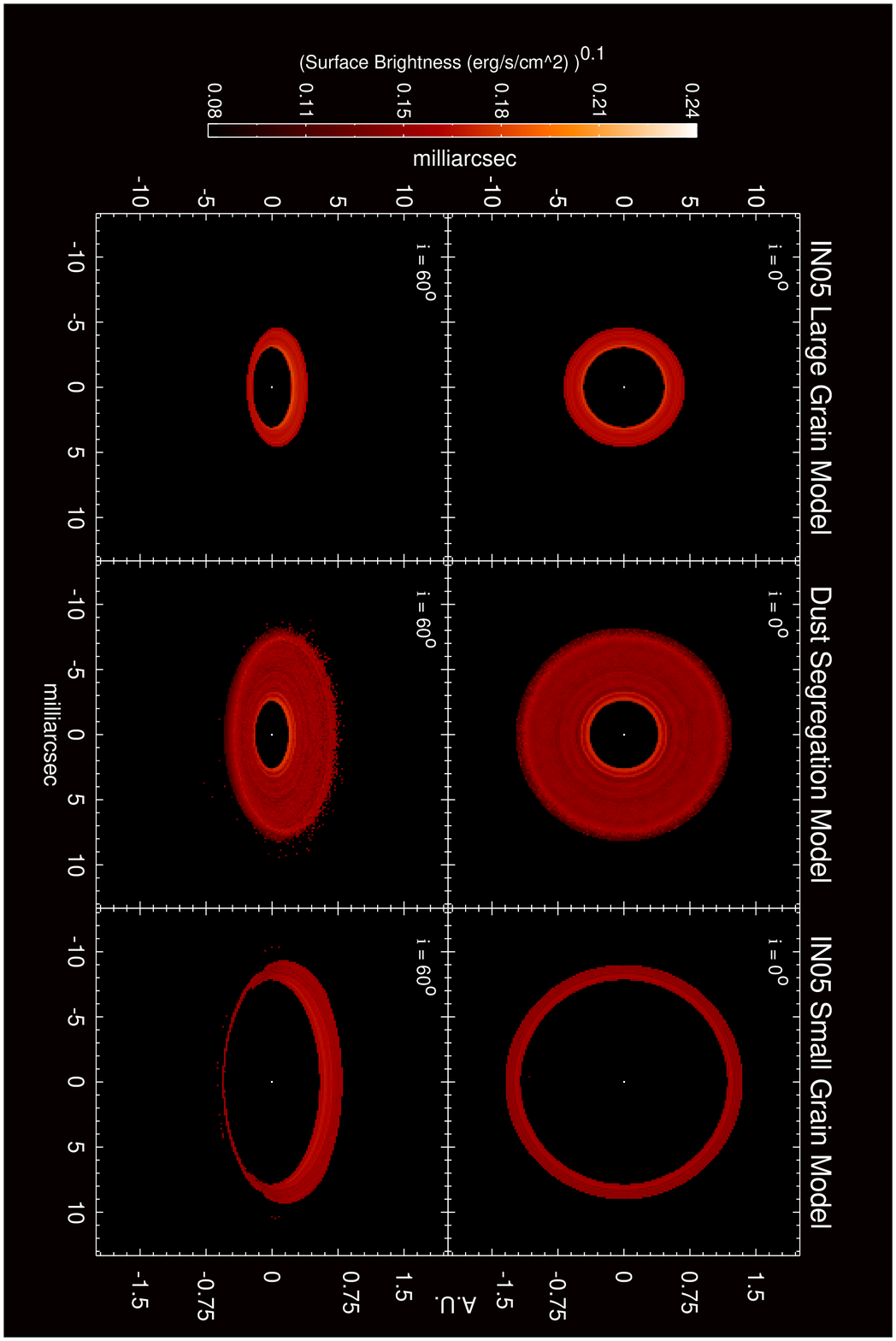}
\caption{\label{fig-tannirkulam-model}Middle column: Model images of the
  first 2-D self-consistent rounded-off rim model
  \citep{Tannirkulam:2007p46688}.  Left and right columns: the IN05 model
  for small (left) and large (right) grains. The Tannirkulam model includes,
  in addition to the dust evaporation physics, also a simple treatment of
  dust growth and sedimentation.}
\end{figure}

The first successful model of a rounded-off rim with 2-D radiative transfer
was presented by \citet{Tannirkulam:2007p46688}
(Fig.~\ref{fig-tannirkulam-model}). They employed AMR, allowing them to
properly follow the very sharp edge of the rounded-off rim. Moreover, they
discovered that grain growth and settling could be another source of
rounding off of the inner rim, because big grains tend to sediment toward
lower $z$ and thus the grains that can survive closer toward the star will
be primarily found closer to the midplane. This effect does not replace the
IN05 effect, but adds to it.

\begin{figure}
\includegraphics[width=28em,angle=90]{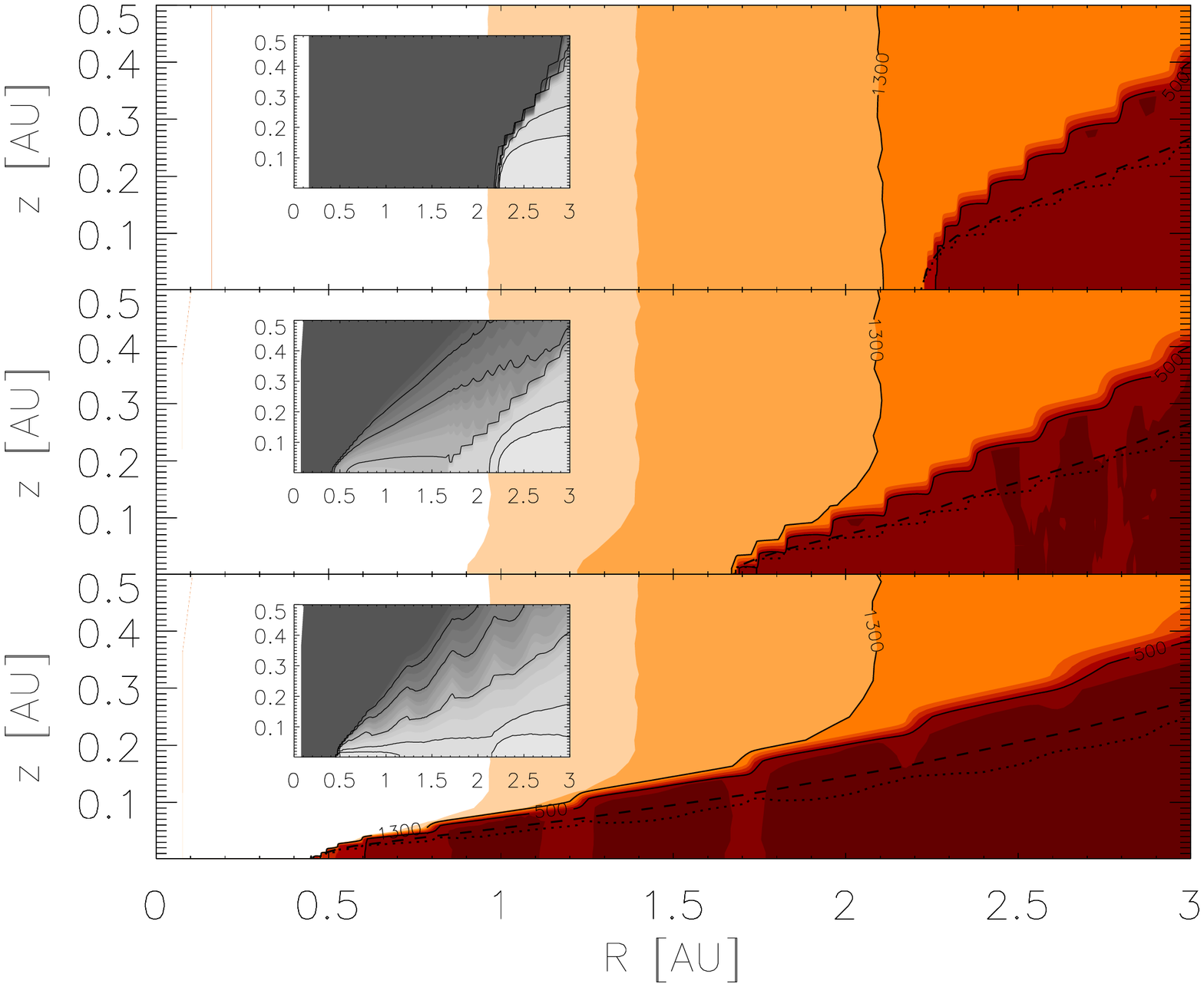}
\caption{\label{fig-kama-model}Model result of \citet{Kama:2009p62149} for
  the temperature structure (red color map) and density structure (grey
  scale insets) of the inner rim of a disk around a star with
  $M_{*}=2.4\,M_{\odot}$, $R_{*}=2.4\,R_{\odot}$, $T_{*}=10000\,$K, for a
  gas surface density of $\Sigma_{\mathrm{g}}=100$ g/cm$^2$. The temperature
  color contours are in steps of 200 K. Density
  contours are a factor $2.7$, $10$, $10^4$ , $10^7$ and $10^{10}$ below the
  maximum. Upper panel: only small 0.1 $\mu$m olivine grains, yielding a
  sharp rim. Middle panel: 1\% of small olivines replaced by 10 $\mu$m
  olivine grains, rest stays small. Lower panel: Like middle panel but now
  replacing 10\%.}
\end{figure}

A recent paper by \citet{Kama:2009p62149} also studied the rounded off rims
with full 2-D radiative transfer, treating the dust evaporation and
condensation physics in more detail than IN05. These models confirm various
aspects of the the IN05 model: the sharp rims for grains with low cooling
efficiency (small $\epsilon$) and the general shape for such rims, the
appearance of translucent rims for high-$\epsilon$ grains and the general
location of the rim. They find that their models that consist of both small
and large grains yield structures that are more like gradually fading disks
toward smaller radii than a sharply defined dust rim, as is shown in
Fig.~\ref{fig-kama-model} (Kama et al.~\citeyear{Kama:2009p62149}; see also
Pontoppidan et al.~\citeyear{Pontoppidan:2007p2909} discussing rims with
mixed dust sizes, though that is a simpler model) .

Clearly, the development of truly 2-D (or even 3-D) self-consistent
radiative transfer models of the dust rim for various dust distributions has
just begun, and much still needs to be done. But this requires very powerful
radiative transfer tools and quite a bit of work to handle them, in
particular with the complex numerical difficulties arising due to the
evaporation and condensation physics.

\subsection{To shadow or not to shadow? The debate on ``puffed up inner rims''}
\label{sec-shadow-or-not}
So far we have discussed the rim structure rather independently from the
rest of the disk behind it. So how do they connect? In the N01/DDN01 papers
it was suggested that the rim may have a thicker geometry than the disk
directly behind it because as a result of the direct frontal irradiation it
is much hotter than the flaring disk behind it that is only irradiated under
a shallow grazing angle (see the paper by Chiang \& Goldreich
\citeyear{chiang:1997} for a nice treatment of flaring protoplanetary disks
and the grazing angle irradiation treatment). The inner rim was thus
expected to be ``puffed-up'' compared to the disk directly behind it. This
would inevitably lead to the rim casting a shadow over at least the inner
few AU of the flaring disk. At larger radii the disk would then ``pop up out
of the shadow'' again and continue outward as a normal flaring disk.  While
DDN01 recognized that due to radial radiative diffusion this shadow would
not be infinitely deep, a relatively strong shadow was nevertheless
predicted.

The 2-D calculations of DD04a confirmed the existence of some form of shadow
covering the inner few AU of the disk, and depending on the distribution of
matter in the disk this shadow could extend even further and in principle
engulf the entire disk (``self-shadowed disks''). The paper showed that
self-shadowed disks have weaker far-infrared (FIR) flux than flared disks,
and confirmed the idea originally put forward by \citet{Meeus:2001p392}
that the observed families of weak and strong FIR Herbig Ae/Be sources
could be explained in this context.

But it was also found that the shadows were much less pronounced than
predicted in the DDN01 model. The radiative diffusion of heat through the
disk in radial direction was much more efficient than predicted by DDN01,
preventing the shadowed region from cooling substantially. Indeed,
\citet{VanBoekel:2005p2865} showed that in the mid-infrared the shadowed
region was merely a slight dip in the intensity profile between
$R_{\mathrm{rim}}$ and $4R_{\mathrm{rim}}$.

A debate ensued whether what was claimed to be a ``shadow'' in the DD04a
paper truly matches the definition of shadow, or whether the weak FIR flux
is merely an effect of the lower surface densities at larger radii in the
``self-shadowed'' models \citep{Wood:2008p42486}. The strictly vertical wall
in DD04a certainly makes the shadow effect stronger than it would be for a
smoother rounded-off rim. Also, even if a shadowing effect takes place, such
a shadow is not ``sharp'' because the top surface of the dust rim is itself
presumably somewhat fuzzy. Finally, due to scattering and radial radiative
diffusion, radiative energy can seep into the shadowed region, preventing it
from cooling drastically. These shadows are therefore certainly not
efficient enough to cause the disk behind the rim to hydrodynamically
``collapse'' toward the midplane for lack of gas pressure support. Even a
small amount of radiative diffusion is enough to heat the midplane of the
shadowed region enough to prevent this. The word ``shadow'' in this context
should therefore be used with care, but can still be useful to explain
certain phenomena in protoplanetary disks.

One argument in favor of some form of self-shadowing taking place in some
objects is the observation of rapid variations in scattered light images of
the outer disk of the Herbig Ae star MWC~275
\citep{Wisniewski:2008p14444}. The time scale of a few years is by far too
short to be explained by hydrodynamic phenomena in the observed outer disk,
which is several hundreds of AU in radius.  Interestingly,
\citet{sitko:2008} report variability in the NIR for this source while the
stellar flux remains steady. This suggests a variable height of the dust rim
which might be linked to variable shadowing of the outer disk. A similar
effect can be observed in the thermal mid- and far-infrared emission from
the outer regions of these disks. For instance, \citet{Juhasz:2007p2911}
have observed variability in the far- infrarared in the source SV Cephei
measured with IRAS on clearly too short time scales to be consistent with
hydrodynamic variability in the far-IR emitting regions. Variable shadowing
appears the only viable explanation \citep{Juhasz:2007p2911}. And recently
\citet{Muzerolle:2009p50951} made multi-epoch Spitzer observations of the T
Tauri star LRLL 31 and found clear differences in the spectra at all
epochs. The mid- to far-IR flux varied by 30\% over time scales as short as
one week. In fact, when the near infrared flux is at its weakest, they found
the mid- and far-infrared flux to be the strongest and vice versa. This
strongly suggests that some rim of variable height casts a shadow over the
outer disk when it is at maximum height, while it leaves the outer disk 
exposed to irradiation if it is at minimum height. One problem is, however,
that \citet{Muzerolle:2009p50951} found that at one epoch there is nearly no
NIR excess at all, suggesting that the inner rim has disappeared completely
at that time. It is unclear how this can have happened.

As mentioned before, however, it is still a matter of debate whether a
hydrostatic dust rim can cast such a shadow. There are, however, other ideas
discussed in the literature on how to create puffed-up inner rims. Already
long before the inner dust rim debate, \citet{Bell:1997p114} found in their
models of strongly accreting disks that the inner disk may be puffed-up due
to the injection of viscous heat deep inside the disk. Since most of the
release of heat happens deep inside the potential well, this puffing up
predominantly affects the inner few AU of the disk, and may cause this inner
disk to cast a shadow of the outer disk. A similar result was found by
\citet{Terquem:2008p50900} for disks with deadzones, where it is not the
heat injection, but the matter storage in the deadzone that puffs the inner
disk up. Conversely, in the models of D'Alessio et al.~the midplane heating
puffs up the disk behind the rim enough to prevent self-shadowing
(N.~Calvet, priv.~comm.).

\subsection{Can the dust rim explain the NIR bump?}
\label{sec-irexcess}
We should now
go back and ask whether the dust inner rim model can indeed explain the
observed NIR bump. The answer is still not conclusive. In contrast to the
simpler N01/DDN01 models, the DD04a models appear to predict an overall
shape of the NIR bump that is flatter (less ``bump like'') and weaker than
what is often observed in Herbig Ae stars. The same was found by Tannirkulam
et al.~(\citeyear{Tannirkulam:2008p41759}) and Isella et
al.~(\citeyear{Isella:2008p49839}). \citet{Vinkovic:2003p48121} were the first
to point out this problem. They made a detailed follow-up study of this
problem in \citet{Vinkovic:2006p52686} and suggest that for sources with a
strong NIR flux, in addition to a disk rim, there must be a second component
such as a spherical envelope that contributes to the NIR flux.

So why did the old models work better in fitting the NIR bump than the more
realistic new ones? This is because they are simple single-temperature
``blackbody wall'' models. From Fig.~\ref{fig-ab-aur-sed} one can see that a
pure blackbody curve fits the bump remarkably well. The more sophisticated
models, however, show that the temperature dispersion in the wall smears out
the flux in the wavelength domain, making the bump flatter and weaker.  To
fix the weakness problem we would need to make the rim geometrically thicker
so that it can capture a larger portion of the stellar light, and thus
become brighter in the NIR (cf.\ Eqs.\ref{eq-lum-covering-fraction},
\ref{eq-def-covering-fraction}). Increasing the pressure scale height of the
disk is difficult: the temperature is more or less fixed (around 1500 K) and
with the radius $R$ and the stellar mass $M_{*}$ the pressure scale height
is given, i.e.\ there is no room for manoeuvring.  Alternatively, by putting
far more matter in the rim, the optical surface moves to a higher elevation
above the midplane. But it may require more density increase than is
realistic. Another idea is to drive a disk-wind that drags along dust and
thus adds an extra dust halo around the disk \citep{Vinkovic:2007p72}.

An alternative explanation, for which there appears to be evidence from NIR
interferometry, is the presence of bright emission from the gas inside the
dust rim (Akeson et al.~\citeyear{Akeson:2005p118}; Monnier et
al.~\citeyear{Monnier:2005p20044}; Tannirkulam et
al.~\citeyear{Tannirkulam:2008p46682}; Isella et
al.~\citeyear{Isella:2008p49839}; Kraus et
al.~\citeyear{Kraus:2008p23127}). The question is, how can one power this
emission without removing power from the dust rim? After all, the gas may
capture stellar radiation, but will also shadow the dust rim behind it. One
suggestion is that active accretion heats up the disk and thus injects the
required additional energy. The other is that if the gas disk extends down
to just a few stellar radii, the finite size of the star allows it to
radiate down onto the disk and thus increase the covering fraction beyond
the simple $\omega=H_s/R$ estimate (see e.g.~Friedjung
\citeyear{Friedjung:1985p361}). But the issue of opacity of the gas inward
of the dust rim is still very much unclear, as we shall see in Section
\ref{sec-inner-gas}.

\subsection{Behind the wall: a dust chemical reactor}
\label{sec-behind-the-wall}
The shape, radius and near-infrared emission of the dust inner rim depends
critically on properties of the dust such as grain sizes and
composition. These are not easy to calculate ab-initio because there are
many processes taking place in the dust inner rim that strongly affect
these. In fact, the region just behind the dust inner rim can be regarded as
a powerful dust chemical reactor. Amorphous dust grains get annealed and
thus acquire crystalline structure (Gail \citeyear{Gail:1998p52714}); iron
may be expelled from the silicates, thus changing the optical properties of
these silicates dramatically (Gail \citeyear{Gail:2004p335}); the iron may
form pure iron grains for which the opacities are hard to calculate; carbon
dust may combust and thus get lost to the gas phase (Gail
\citeyear{Gail:2002p23391}).

In addition to these chemical processes, the high temperatures also provide
for interesting physics to occur. For instance, dust particles that collide
at these high temperatures may stick and melt together due to processes such
as sintering and/or eutectic melting \citep[see
e.g.][]{Blum:2008p29949}. This may aid the growth of grains in these rims,
meaning that maybe these rims are rich in large molten-together grain
clusters. While large grains or
grain clusters can exist closer to the star than small grains, if they
evaporate nevertheless (for instance due to a temporary fluctuation in the
brightness of the star due to some transient accretion event), they will not
be able to recondense at those radii because their growth has to pass
through a phase of small grains, and these small grains would only survive
at larger radii. Large grains can then only re-appear in these regions if
they are transported inward from larger radii.

All of these effects have relevance also for the rest of the disk. The
thermally processed material that is created here in this hot ``oven'' may
be partially transported outward to the planet- and comet-forming region of
the disk, for instance through radial mixing (e.g.\ Gail 
\citeyear{Gail:2001p23416}; Bockel{\'e}e-Morvan et
al.~\citeyear{BockeleeMorvan:2002p25424}). There is evidence for this from
ingredients found in the material brought back from comet Wild-2
\citep{Zolensky:2006p26057} as well as from infrared observations (e.g.~van
Boekel \citeyear{VanBoekel:2004p298}). The study of the dust inner rim is
thus also the study of an important preprocessing factory of protoplanetary
building material. Clearly there is still a lot to be investigated in this
field.

\section{Gas inward of the dust rim}
\label{sec-inner-gas}
One of the main assumptions in the inner dust rim model is that the gas
inward of the rim is optically thin, so that the star can freely illuminate
the dust rim. However, this assumption is rather crude. Already soon after
the first dust rim models came out, \citet{Muzerolle:2004p50} made an
investigation of the gas optical depth inward of the dust rim, and they
found that for low accretion rates the inner gas is sufficiently transparent
for the dust rim to be appreciably illumated by the star. For high accretion
rates ($\dot M\gtrsim 10^{-8}M_\odot/$year) they found that the inner gas
disk becomes optically thick. However, with recent progress in observational
studies of the gas inward of the rim (Section \ref{sec-gas-obs}) this topic
may need to be revisited. In this section we shall discuss current
understanding of the dust-free inner disk, thereby moving our attention
closer toward the star. We will follow largely the study of
\citet{Muzerolle:2004p50}.

\subsection{Gas opacities}
\begin{figure}
\includegraphics[width=35em]{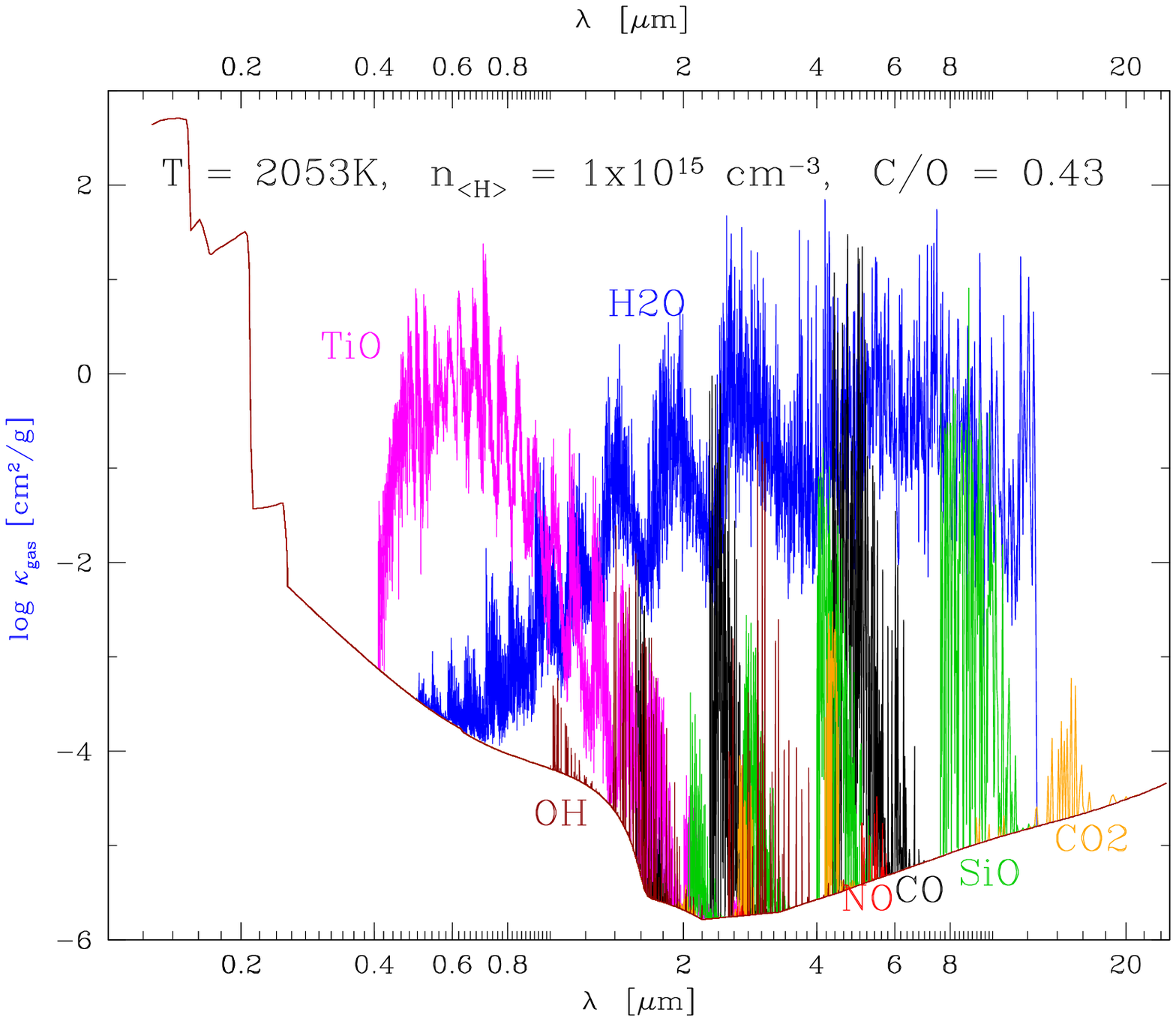}
\caption{\label{fig-gas-opacities}An example of opacities of the gas at
  conditions typical for the inner dust-free region of the disk. The
  abundances of the various sources of opacity (TiO, H$_2$O being the
  dominant line absorption agents) are calculated using equilibrium
  chemistry at the temperature of 2053 K. No photodissociation of the
  molecules is included which might substantially reduce the strength of the
  molecular line contribution to the opacity. From Ch.\ Helling priv comm.,
  used with permission \citep[see][for details on the
  computations]{Helling:2009p49532}.}
\end{figure}

The first thing to do is to get a grasp of the gas opacities at temperatures
between $T_{*}$ and $T_{\mathrm{rim}}$ (or even slightly below that) and
densities appropriate for our purpose. One may be tempted to use one of the
publically available tables of Rosseland- or Planck-mean gas opacities
available on the web. But such opacities are valid only in optically thick
media, or in media of medium optical depth ($\tau\sim 1$) that are not
irradiated by an external source. In our case we expect that neither of
these conditions are met. At low accretion rates the gas will be at least
partially optically thin, and it will be strongly irradiated by the light of
the star, which has a color temperature much in excess of the gas kinetic
temperature. We are therefore forced to use some form of frequency-dependent
opacities. In most cases the gas is at too low temperatures for strong
continuum opacity sources such as H$^{-}$ to play a substantial role, except
in the very tenuous surface layers of the disk where temperatures may be
very high \citep{Glassgold:2004p54989}.  Instead, depending on the local
chemistry, we may be faced with a zoo of billions of molecular and atomic
lines. An example of such a zoo is shown in Fig.~\ref{fig-gas-opacities},
which was calculated for a temperature of about 2000 K and a density of
$\rho_{\mathrm{gas}}\simeq 4\times 10^{-9}$ g/cm$^3$, and based on chemical
equilibrium abundances of molecular and atomic species.

Whether the actual gas opacity of the dust-free inner disk is as rich in
molecular lines as the opacity shown in Fig.~\ref{fig-gas-opacities} is,
however, not clear. Molecules are fragile and are collisionally destroyed at
temperatures much in excess of 2500 K. Moreover, UV photons from the star
and from the accretion shocks near the stellar surface can photodissociate
these molecules even if they are well below the thermal dissociation
temperature. The destruction of molecules, if it proceeds all the way to
atoms, would clearly reduce the opacity dramatically. For H$_2$O the problem
of photodissociation in dust-depleted surface layers of disks has been
studied by \citet{Bethell:2009p50175}. They found that water is quite
efficient in shielding itself from the destructive UV photons of the
star. And \citet{Glassgold:2009} suggest that the more the dust is removed
from the disk, the stronger the observed infrared H$_2$O lines should
become. But none of these models are directly applicable to the very inner
dust-free disk, and their physics and chemistry is still far from
complete. Clearly a lot still has to be investigated before a good picture
emerges of the composition and opacity of the gas inward of the dust rim.

If we, for now, take the opacities of Fig.~\ref{fig-gas-opacities} at face
value, then one is faced with the question how to deal with the complexity
of it, involving billions of molecular lines. While the opacity can be very
high at the line centers, there are opacity valleys between lines, so
radiation can seep through these ``leaks''. And there are regions with few
or no lines where, because there is (by assumption) no dust inside of the
dust inner rim, there is only a weak gas continuum opacity present. This is
particularly so for the region between about 0.2 and 0.4 $\mu$m.  Stellar
radiation is allowed to pass virtually unhindered through such a
``window''. There is no ultimate answer to the question how to handle this
complexity. But by assuming that all abundances and excitations of the
molecules are in local thermodynamic equilibrium (LTE) and by using
techniques such as ``opacity sampling'' \citep{Ekberg:1986p57820} or
multi-group methods (Mihalas \& Mihalas \citeyear{Mihalas:1984p52278}),
progress can be made.

Another method, employed by \citet{Muzerolle:2004p50}, is to treat the large
opacity gap between 0.2 and 0.4 $\mu$m (Fig.~\ref{fig-gas-opacities})
separately from the rest, and treat the rest using special-purpose mean
opacities, constructed particularly for the problem at hand. Let us now have
closer look at that paper.

\subsection{Structure of the dust-free gas inner disk}
\citet[henceforth MDCH04]{Muzerolle:2004p50} used the basic framework of
disk structure of the D'Alessio disk models (D'Alessio et
al.~\citeyear{DAlessio:1998p71}; Calvet et al.~\citeyear{Calvet:1991p8490})
and replaced the dust opacities with the appropriate gas mean opacities to
obtain a model of this part of the disk. To overcome the problems with the
use of mean opacities, they constructed a specially tailored mean opacity
which acts as a Planck mean, but takes into account the opacity gaps between
the lines in the $\lambda>0.45\mu$m regime. For $0.2<\lambda<0.45\mu$m they
assume appropriately weak continuum opacity. In this section we will
construct a simplified version of that model as an illustration of the gas
inner disk issue.

Following MDCH04, we use a razor-thin disk approach, so that the irradiation
of the disk by the star occurs solely due to the finite size of the star,
allowing it to shine down on the disk. The average angle under which this
radiation then hits the disk is
\begin{equation}
\varphi(R) \simeq \frac{4}{3\pi}\;\frac{R_{*}}{R}
\end{equation}
We also assume that only the upper half of the star surface is able to
radiate onto the upper surface layers of the disk (MDCH04 put this
additional factor of 1/2 inside their definition of $\varphi(R)$). MDCH04
found that at these very shallow angles the stellar radiation longward of
$0.45\mu$m will be fully absorbed by the surface of the gas disk. They also
found that for small enough accretion rate the temperature of this surface
layer is always around 2000 K. Now following the reasoning of
\citet{chiang:1997} this layer re-emits the absorbed energy at its own
temperature (2000 K), producing emission in the NIR. Half of this radiation
is lost upward, the other half is radiated downward into the disk.  At these
wavelengths the opacity of the gas hovers between $\kappa_\nu\sim 10^{-2}$
and $10^{1}$ cm$^2$/g (cf.\ Fig.~\ref{fig-gas-opacities}, but keep in mind
the caveats). Below we will verify whether the gas disk is optically thick
in vertical direction at all these wavelengths; for now we make this
assumption and find, by balancing the downward flux
$F_{\mathrm{down}}=(1/4)f\varphi(R)L_{*}/(4\pi R^2)$ (with $f$ explained
below and the factor $1/4$ accounting for the two factors $1/2$ discussed 
above) with upward the cooling flux $F_{\mathrm{up}}=\sigma
T_{\mathrm{mid}}^4$ that the midplane temperature is
\begin{equation}\label{eq-tmid-inner-gas-disk}
T_{\mathrm{mid}} = T_{*} \sqrt{\frac{R_{*}}{R}}\;(\tfrac{1}{4}f\varphi(R))^{1/4}
=\left(\frac{R_{*}^3f}{3\pi}\right)^{1/4}T_{*}\;R^{-3/4}
\end{equation}
where again we assumed $L_{*}=4\pi R_{*}^2\sigma T_{*}^4$ for simplicity.
The factor $f$ tells how much of the stellar radiation is longward of 0.45
$\mu$m. Let us take this factor to be $1/2$.
%
%
This estimate does not include
the heating effect by viscous accretion in the disk. According to MDCH04's
solutions, however, this only kicks in for accretion rates $\dot M\gtrsim
3\times 10^{-8}M_{\odot}/$yr, so we will ignore this here. One can now plot
the above result and compare to Fig.~9 of MDCH04 to find that the match is
not bad for the cases of $\dot M\lesssim 10^{-8}M_{\odot}/$yr.

With this estimate of the temperatures of the midplane and the surface
layers of the gaseous dust-free inner disk, we can calculate the surface
density in the disk. We need standard Shakura-Sunyaev-type accretion disk
theory (see e.g.\ Hartmann \citeyear{hartmann:2009}), which states that the
accretion rate is $\dot M=3\pi \Sigma_{\mathrm{gas}}\nu_t$, with
$\Sigma_{\mathrm{gas}}$ being the gas surface density and $\nu_t=\alpha k
T_{\mathrm{mid}}/\mu_g \Omega_K$ being the turbulent viscosity. The constant
$\alpha$ is the ``turbulent viscosity coefficient'' which we take to be the
standard value of $\alpha=0.01$. This leads to the following powerlaw
expression for the surface density of the gas:
\begin{equation}
\Sigma_{\mathrm{gas}}(R) = C\;\dot M\;
  R^{-3/4}\qquad \mathrm{with}\quad 
 C\equiv \frac{\mu\sqrt{GM_{*}}}{(3\pi)^{3/4}R_{*}^{3/4}\alpha kT_{*} f^{1/4}}
\end{equation}
(valid for $\dot M\lesssim 10^{-8}M_{\odot}/$yr). Putting in our standard
numbers we obtain: $\Sigma_{\mathrm{gas}}(R)=350\,(R/\mathrm{AU})^{-3/4}$
g/cm$^2$ for $\dot M=10^{-8}M_{\odot}/$yr (scaling linearly with $\dot
M$). With the assumed opacities this validates the previous assumption that
the disk is optically thick to the surface layer radiation for this
accretion rate.

We can also estimate the pressure scale height $H_{\mathrm{p}}(R)$, which
(see Eq.~\ref{eq-hp-definition}) amounts to
$H_{\mathrm{p}}/R=0.0167\,(R/\mathrm{AU})^{1/8}$ for our example
model. Note that the ratio $H_{\mathrm{p}}/R$ (the disk opening angle) is
nearly constant, at around 0.014. Unfortunately this result shows that the
razor-thin disk assumption is not entirely justified, since 0.014$\times
0.5\mathrm{AU}=0.6\,R_{*}$, meaning that at 0.5 AU the disk's pressure scale
height is already half the star radius, and likely the surface height (where
the disk becomes optically thin to stellar light incident at a small angle
$\varphi$) substantially above that. So the model has to be refined, but as
a rough estimate, it suffices.

\begin{figure}
\includegraphics[width=35em]{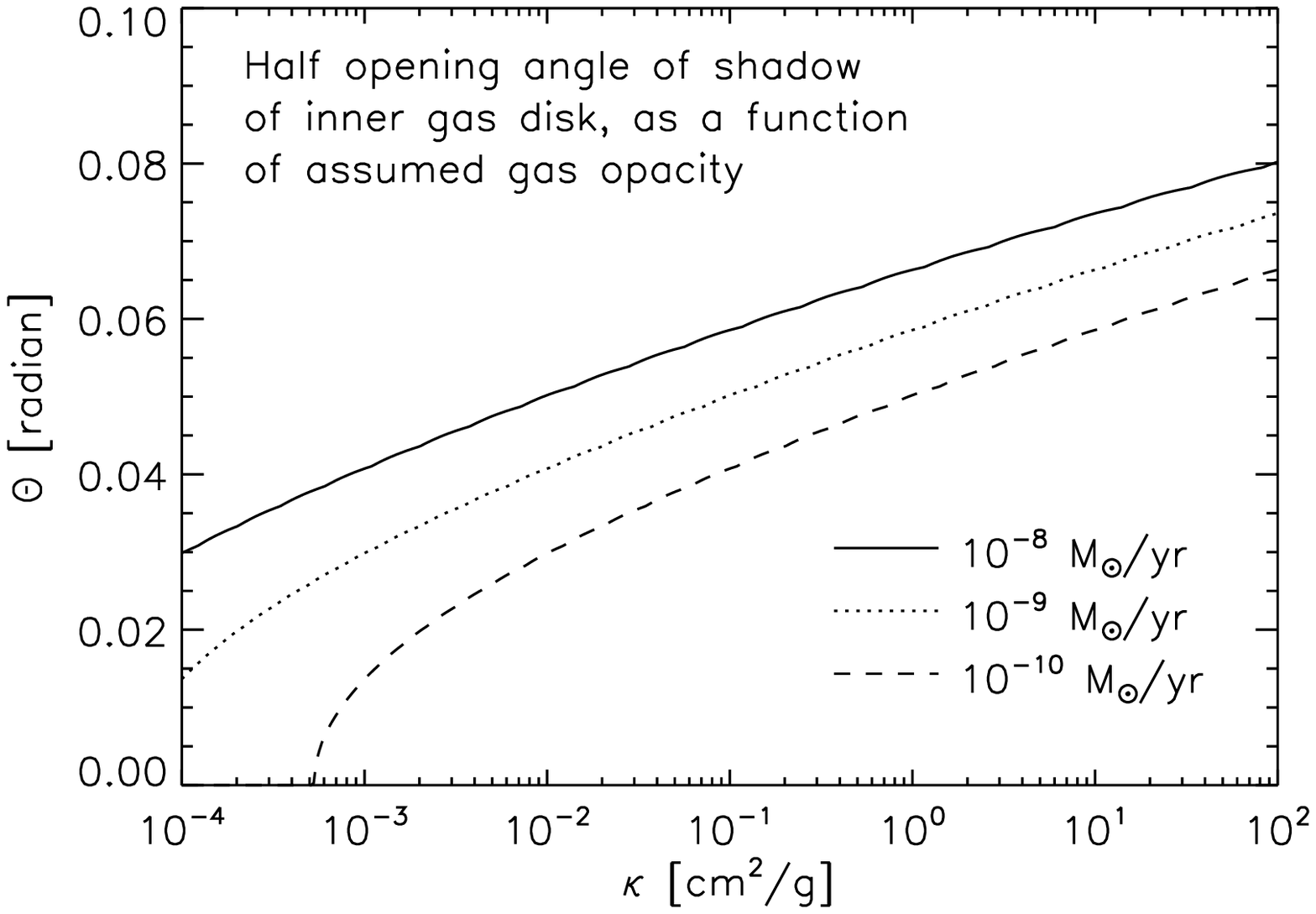}
\caption{\label{fig-theta-vs-kappa}The half-opening-angle of the 
shadow cast by the gaseous dust-free inner disk as a function of
gas opacity $\kappa$, for our standard example star and for 
three different accretion rates. The inner radius was assumed 
to be at 0.03 AU and the outer radius at 0.5 AU. See text for details.}
\end{figure}

Next we can calculate the opening angle of the shadow that the gas inner
disk casts on the dust inner rim. For this we need the gas density, which
follows from the application of Eq.~(\ref{eq-gaussian-density}) to the gas
disk.  We can then numerically integrate radially outward to the dust rim,
assuming for simplicity now that the star is a point source, at various
angles $\Theta$ with respect to the midplane ($\Theta=0$ meaning through the
midplane). This gives the radial column density of gas. Multiplying this
with some guess of the gas opacity $\kappa$ gives the optical depth for that
$\kappa$, which is of course depending on $\nu$. We can now plot, as a
function of $\kappa$, at which $\Theta$ this optical depth drops below
unity. This gives the half-opening-angle of the shadow cast by the gas disk
for that particular value of $\kappa$. This plot is shown, for our model, in
Fig.~\ref{fig-theta-vs-kappa}.

The angle $\Theta$ can be interpreted as follows. If the dust inner rim has
a vertical optically thick surface height of about 0.2 AU at a distance of
1.5 AU (as we take from our example in Fig.~\ref{fig-rounded-rim-model})
then a shadow of $\tan\Theta=0.2/1.5=0.133$ would mean that the gas disk
completely shadows the expected dust inner rim. This would mean that the
dust rim model as it stands would not work. It will be cooler, and thus have
a less high vertical extent, and it will emit little NIR flux. From
Fig.~\ref{fig-theta-vs-kappa} we get, however, that the shadow half-opening
angle is at most $\Theta=0.08$, for relatively high gas opacity as is
present in some of the lines of water and TiO. However, for the opacity gap
between 0.2 and 0.4 $\mu$m where, say, $\kappa\sim \mathrm{few}\times
10^{-3}$ cm$^{2}$/g, the shadow is between $\Theta\simeq 0.02$ and
$\Theta\simeq 0.05$. What we can conclude here is that the very midplane of
the gas inner disk will likely cast some form of shadow on the inner dust
rim. However, this shadow is likely not geometrically thick enough to cover
the entire rim (MDCH04), and in particular in the opacity gap between 0.2
and 0.4 $\mu$m it is a relatively thin shadow. This situation is
pictographically shown in Fig.~\ref{fig-picto-inner-wall}. Moreover, the
estimate in Fig.~\ref{fig-theta-vs-kappa} is presumably overly pessimistic,
because due to the finite stellar size, much of the radiation will simply
move over the gas inner disk instead of through it. And finally, if the
molecular abundances are substantially suppressed due to
(photo-)dissociation, perhaps the shadow becomes weaker still.

\section{Probing the inner dust-free disk with gas line observations}
\label{sec-gas-obs}
After some substantial theoretical modeling, let us return to observations,
and focus in particular on what we can learn from high-spectral-resolution
optical and near-infrared observations.

\subsection{The search for molecules in the inner dust-free disk}
Due to the absense of dust in the very inner disk regions, and the tentative
theoretical expectation of substantial molecular content of the gas there,
one may suspect that these regions are potent molecular line emitters.
Interestingly, though, so far molecular line emission from these regions is
only occasionally seen. In part this may be because of current observational
limitations. But there is some evidence that there may be a true deficit of
molecules in the dust-free inner disk.

\citet{najita:2007} recently reviewed the major results from high spectral
resolution observations on YSO disk using large aperture telescopes,
including a number of studies that impact the inner AU of disks.  The high
orbital velocities of the material in the inner disk as well as the high
excitation temperatures of ro-vibrational transitions of molecules such as
CO and H$_2$O can be used to probe the dynamical and chemical compositions
of these inner disks in a powerful way.  CO fundamental lines are commonly
observed in YSO disks and likely form in the surface layers of the disk over
a large range of radii \citep[from $\lesssim 0.1$ AU out to $\sim 2$ AU for
solar mass young stars][]{najita:2007}.  Interestingly, the CO overtone
lines, which are collisionally populated by hot gas at $>$1000K, are much
more rare: only a few percent of T Tauri stars surveyed. In the rare cases
where CO overtone lines are strong, often hot water bands are also seen
\citep{carr:2004,najita:2009,Thi:2005p47961}. CO overtone emission probably
originates from the very inner regions \citep[$\sim 0.05$ AU to $0.3$ AU for
solar mass young stars][]{najita:2007}. Since $R_{\mathrm{rim}}$ for these
stars is expected to lie roughly near $0.1$ AU, the rarity of overtone
emission seems to suggest that much of the molecular content is destroyed in
the dust-free inner gas disk. However, it should be kept in mind that
overtone emission requires not only a higher temperature, but also a higher
column of CO than fundamental emission to be detectable. So it could equally
well mean that most of the gas in this dust-free inner disk is cooler than
expected, and that the total column of hot gas is too low for detectable
overtone emission.

A recent advance in NIR interferometry is ``spectro-interferometry'', in
which the interferometric signal (visibility) can be measured as a function
of wavelength. If the NIR flux of an object contains an interesting
gas-phase line, then by determining the change in visibility over this line
the differences in spatial scale of the emission of the gas line and the
underlying continuum can be established.  In some cases one can even
retrieve information for different velocity components. This technique was
pioneered for H-$\alpha$ observations in the visible for classical Be stars
\citep{mourard:1989}, and was first applied to Br-$\gamma$ emission
measurements of YSOs by \citet{malbet:2007} for the wind-dominated Herbig-Be
star MWC 297 and in the normal Herbig Ae star HD~104237 by
\citet{tatulli:2007}.  The latter result was shocking: the Br-$\gamma$ was
not coming from the accretion flows as expected, but was arising from an
extended region just inside the dust evaporation radius.  Subsequent
measurements on larger datasets \citep{kraus:2008, eisner:2009} find a
diversity of size scales for the Br-$\gamma$ emittion, from point-like to
extended. The spatial extent can be both larger and smaller than that of the
continuum. A comprehensive understanding has yet to emerge but may be
related to the strength of the inner disk wind in Herbig Ae/Be stars.  The
distribution of Br-$\gamma$ emission is possibly linked to the hot inner
disk emission observed in the continuum discussed in \S\ref{sec-inner-gas},
since in some cases they are co-spatial.

Br-$\gamma$ is typically the strongest gas signal seen with NIR
spectro-interferometry. For the rest the visibilities as a function of
wavelength have so far been observed to be rather ``continuum like''.
Sometimes a hint of molecular emission from the hot inner disk is observed,
such as the CO overtone band heads in the spectrum of RW Aur
\citep{eisner:2009} or the weak water feature tentatively found in the inner
disk of Herbig Ae star MWC~480 \citep{eisner:2007b}. But these signals, if
they are real, are very weak. In fact, high resolution spectroscopic
single-telescope K-band observations of some of these same sources
\citep{najita:2009} do not show any strong molecular lines, no CO overtone
nor emission from water or any other molecules, though CO fundamental
emission is detected for MWC 480 \citep{Blake:2004p62015}. A comparison to
the molecular line forest expected from Fig.~\ref{fig-gas-opacities}
suggests that maybe these inner gaseous disks are not nearly as rich in
molecules as we expected. In fact, as pointed out by Benisty et al.~(in
press) even gas continuum opacities such as H$^{-}$ opacity seem to be in
conflict with the data. This poses the question: what is causing the smooth
continuum emission from inside the dust rim? Currently this seems to be an
unsolved question.

It is really a mystery why YSO disks do not show stronger emission from
molecules in the dust-free inner disk, given that the molecular emission is
regularly seen in the surface layers of the disk in the dusty regions of
these disks \citep{carr:2008,mandell:2008,salyk:2008,bitner:2008}.
But there does seem to be evidence for molecules right within the dust rim
itself, where perhaps the dust may play a role in protecting these molecules
from the photodissociating radiation. A nice example of such an observation
is the work by \citet{Lahuis:2006p2917}. They reported the detection of hot
organic molecules in absorption against the infrared continuum with
Spitzer-IRS in the object IRS 46. They argue that these absorption lines are
observed because the line of sight toward the hot inner dust rim is partly
blocked by the back side of this rim. If confirmed, this would show that
molecules are at least present in the dusty rim itself. Interestingly,
variability of these absorption lines is observed (Lahuis in prep),
suggesting that the rim may cause variable levels of extinction for some
reason. This suggests some link to the idea, suggested by
\citet{natta:2001,Dullemond:2003p2872,Pontoppidan:2007p2909}, that
variability in the rim, as seen under a strong inclination, may in fact
be the cause of UX Orionis-type short-timescale extinction events seen in
some sources.

\subsection{Probing the dynamics of the inner gas disk}
Apart from looking for the presence of molecules, the study of molecular
lines from the inner disk also allows us to study the dynamics of these
inner regions. A particularly powerful technique is the emerging specialty
of spectro-astrometry in the infrared, where an AO-corrected large-aperture
telescope is used to measure centroid shifts as a function of spectral
channel.  Breakthrough results by \citet{pontoppidan:2008} show Keplerian
motion in the fundamental CO band with evidence for strong departures of
symmetry in some sources, demonstrating that even single telescopes can
access information at the sub-milli-arcsecond level in some cases.  Also
spectro-astrometry can uncover evidence of binarity when emission lines do
not trace continuum \citep{baines:2006}.

Another technique that has proved useful for probing the kinematic and
spatial properties of the innermost gas disk is spectro-polarimetry.  By
measuring the polarization of the continuum along with the polarization of
the H-$\alpha$ emission line at multiple velocity channels, one can deduce
the geometry and rotation properties of the innermost gas (within a few
stellar radii) in YSOs \citep[e.g.,][]{vink:2002,vink:2005}.
\citet{vink:2005} find evidence that the innermost gas in T~Tauri and Herbig
stars is oriented in a rotating disk aligned with the outer disk seen by
other techniques (e.g., millimeter or scattered light) and that the Herbig
Ae disks are significantly optically-thinner than the T Tauri ones.  The
potential for this technique has not yet been fully realized and we hope to see
continued progress \citep[e.g., see recent work of][]{harrington:2007,
Mottram:2007p48511}.

\section{Summary and outlook}
In this review we have shown that the inner $\sim$AU of protoplanetary disks
is an area of rich physics. It is the region where dust is chemically
processed and evaporated. We have seen that the early naive picture of a
definite ``evaporation radius'' separating the dusty outer disk from a
dust-free inner disk is too simple. Complex interplay between
multi-dimensional radiative transfer, dust chemistry and dust evaporation,
together with insufficiently well understood gas opacities, yield a picture
in which the transition from the dusty outer disk to the dust-free inner
disk is far more gradual and subtle than previously thought. The complexity
of the physics involved is so great, that even after almost a decade of
research no definite model is in sight. And in this review we have not even
had the chance to touch upon additional highly important and complex
inner-disk phenomena such as magnetospheric accretion and episodic/unstable
accretion. This means that the inner regions of protoplanetary disks are
still a fertile ground for exciting research in the years to come, both
observationally and theoretically. 

We have already touched upon a number of issues that we consider not yet
resolved. But it may be useful to give here a summary of these -- and a 
number of not-yet-mentioned -- issues:
\begin{itemize}
\item {\em Radiative transfer:} Does/can the rim cast a shadow over the disk
  (``self-shadowing'')?  How do radiation pressure and possibly
  photophoresis affect the distribution of dust in the inner rim region?  Is
  the rim solely responsible for the NIR bump, or are there other sources as
  well, and what is the nature of these sources? Can disk winds contribute
  to the NIR bump?
\item {\em Dust composition:} What is the interplay between chemistry,
  condensation, evaporation, turbulent mixing and drift of dust grains in
  the dust inner rim?  And consequently, what is the composition of the dust
  in the rim?  Does the dust mineralogical processing in the rim affect
  planet formation and/or the composition of planets and planetary debris?
\item {\em Gas inward of the 'dust rim':} What is the cause of the observed
  emission inside the dust rim? What is the dominant source of opacity in
  this region: Molecular lines, atomic lines, gas continuum or dust after
  all? Why are only few molecular lines from these very inner regions
  observed so far? Is the gas inward of the dust rim optically thin or
  thick? Can the gas ``protect'' the dust by its shadow, thus allowing dust
  to survive closer to the star? What the emission mechanism for the
  Br-$\gamma$ line in Herbigs and T~Tauris? Why does the spatial extent of
  this feature show such diversity?
\item {\em Dynamical behavior:} What is the (hydro)dynamic behavior of the
  rim? Is it stable? How does the internal heating near the midplane of the
  disk by active accretion affect the structure of the inner disk?  How do
  we understand the reports of short-term variability of the inner disk in
  some sources \citep{sitko:2008,tuthill:2002,millan-gabet:2006}?  Any
  relation to production of micro-jets \citep{devine:2000}?
\end{itemize}
These are just a number of issues, among many more. With advances in optical
and infrared interferometry, with techniques such as spectropolarimetry and
spectroastrometry, in particular on ELTs, and with more focus on variability
of all measured observables, and last but not least, with continued improvement
of theoretical modeling, we believe that in the coming 10 years we will likely
find the answers to many of the above questions.

\vspace{1em}

We thank H-P.~Gail, C.~Dominik, H.~Linz, R.~van Boekel, U.~Jorgensen,
C.~Helling, Th.~Henning, M.~Min., R.~Millan-Gabet, J.~Najita, Th.~Preibisch,
N.~Calvet, S.~Krauss, J.~Bouwman, L.~Waters and A.~Natta for many
interesting and useful discussions and comments that have helped us a lot.

\end{document}